\documentclass[preprint]{aastex62}

\usepackage[inline]{enumitem}

\submitjournal{Icarus}

\shorttitle{}
\shortauthors{Ribeiro et al.}

\begin{document}

\title[Reassessing the origin and evolution of Short-Period Comets in the Planet-9 Scenario]{Reassessing the origin and evolution of Ecliptic Comets in the Planet-9 Scenario}

\correspondingauthor{Rafael Ribeiro de Sousa}
\email{r.sousa@unesp.br}

\author[0000-0002-0786-7307]{Rafael Ribeiro de Sousa}
\affiliation{S\~ao Paulo State University, UNESP,Campus of Guaratinguet\'a, Av. Dr. Ariberto Pereira da Cunha, 333 - Pedregulho, Guaratinguet\'a - SP, 12516-410, Brazil}

\author{Andre Izidoro}
\affil{Department of Earth, Environmental and Planetary Sciences, 6100 Main MS 126,  Rice University, Houston, TX 77005, USA}

\author{Alessandro Morbidelli}
\affil{Laboratoire Lagrange, UMR7293, Université Côte d’Azur, CNRS, Observatoire de la Côte d’Azur, Boulevard de l’Observatoire, 06304 Nice Cedex 4, France}

\author{David Nesvorny}
\affil{Southwest Research Institute, 1050 Walnut St. Suite 300, Boulder, CO 80302, USA}

\author{Othon Cabo Winter}
\affil{S\~ao Paulo State University, UNESP,Campus of Guaratinguet\'a, Av. Dr. Ariberto Pereira da Cunha, 333 - Pedregulho, Guaratinguet\'a - SP, 12516-410, Brazil}

\begin{abstract}
\small
A group of newly observed extreme trans-Neptunian objects show an unexpected level of orbital confinement, characterized by an approximate alignment of the  orbital angular momentum vectors and apsidal lines. It has been proposed that a yet undiscovered giant  planet (named {\it Planet-9}) exists in the outer parts of the solar system and is causing this clustering.  Initial studies suggested that Planet-9 could be as massive as 15$M_{\oplus}$. In this mass range, however, this planet tends to strongly interact with scattered disk objects (SDOS; $50 < a < 1000$~au) and influence the dynamics and the orbits of a population of short period comets known as ecliptic comets. The outcome of this interaction is a population of ecliptic comets with orbital inclinations broadly inconsistent with observations.  In this work, we model the formation and long-term dynamical evolution of trans-Neptunian object populations and Oort cloud during the solar system dynamical instability phase considering a revised set of mass and orbital parameters for Planet-9. In our simulation, Planet-9 is assumed to have a mass of $m_9 \sim 7.5 M_{\oplus}$, a moderately inclined orbit with $i_9 \sim 20$ deg,  semi-major axis $a_9 \sim 600$ au, and orbital eccentricity of $e_9 \sim 0.3$. Our results show that a relatively less massive Planet-9 is broadly consistent with the inclination distribution of trans-Neptunian objects and the observed number of  ecliptic comets ($D>10$ km ) in the solar system. Furthermore, our results indicate that under the influence of Planet-9, distant Kuiper belt objects with \( 40 < q < 100 \, \mathrm{au} \) and \( 200 < a < 500 \, \mathrm{au} \) that are significantly inclined, are more likely to be apsidally aligned with the planet rather than anti-aligned, with an anti-aligned-to-aligned population ratio of approximately 0.5-0.7. Objects within this semi-major axis and perihelion range and with orbital inclinations lower than $\lesssim$ 20 deg (comparable to that assumed for Planet-9), however, exhibit significant apsidal anti-alignment. Within this low-inclination subset, the ratio of anti-aligned to aligned populations is approximately 2-4. These findings provide a novel observational direction that could help refine the search for this putative planet.

\end{abstract}

\section{Introduction}

The migration history of the giant planets in the early Solar System played a central role in shaping the dynamical structure of the present-day Solar System. The so-called \textit{Nice model} proposes that the giant planets experienced a dynamical instability  triggered by interaction with a primordial planetesimal disk existing in the outer solar system \citep{Tsiganisetal2005,Levison2011,NesvornyMorby2012}. There are several lines of evidence supporting this model. The giant planet instability scenario is consistent with the Trojans asteroids of Jupiter and Neptune \citep{Morbidellietal2005,Nesvornyetal2013,GomesNesvorny2016,Deiennoetal2017}, the irregular satellites of the giant planets
\citep{Nesvorny2007,Nesvornyetal2014}, the Kuiper belt's orbital structure  \citep{Nesvorny2015a,Nesvorny2015b,Nesvorny2016a,Gomesetl2018,Nesvorny2018}, and trans-Neptunian scattered disk objects and Oort Cloud as the source of comets in the Solar System \citep[e.g.][]{Nesvornyetal2017}.

While the original Nice Model and its variations have been successful in explaining several outer
Solar System constraints,  recent observations of the distant trans-Neptunian objects (TNOs, objects with semi-major axis beyond the orbit of Neptune) show a subset of objects with an unexpected orbital confinement \citep{2014Natur.507..471T,2016AJ....151...22B,2021ApJ...910L..20B}. This observed alignment is mostly  characterized by an approximate clustering of the orbital angular momentum vectors and apsidal lines~\citep{2014Natur.507..471T,BB2019}. Clustered objects show orbits with semi-major axis overall larger than 250 au. This clustering is intriguing because objects orbiting at such distances from the Sun should show no preferential orbital orientation, unless a powerful physical mechanism is acting to maintain it over long timescales.

It has been proposed that the observed TNOs' clustering  is the dynamical signature of a yet undiscovered, massive and distant planet orbiting in the very distant solar system. This planet has been refereed as Planet-9. Planet-9 mass has been initially proposed  to be $\gtrsim 10M_{\oplus}$ \citep{2016AJ....151...22B} but it has been more recently constrained to be about $\sim$5-10$M_{\oplus}$~\citep{BB2019,2021ApJ...910L..20B}.

It not entirely clear how such a massive Planet-9 can affect the formation and the dynamical evolution of the cometary reservoirs, and whether it is consistent with the  current distribution of observed comets or not. In this regard, short period comets (P$<$20 yr) are a particularly strong constraint. Their population is relatively well characterized from observations and they can be effectively used to test this hypothesis~\citep{Nesvornyetal2017} .

Recent work by ~\cite{Nesvornyetal2017} shows that Planet-9 tends to increase the orbital inclination dispersion of the trans-Neptunian scattered disk objects ($50<a\leq 1000$ ~au, $q>40$~au)  which reflects in the  inclination distribution of ecliptic comets \citep{Nesvornyetal2017}. Ecliptic comets (ECs) are a sub-group of short period comets (P$<$20 yr) with Tisserand parameter values extending from 2 to 3. The inclination distribution of ECs in the presence of Planet-9 found by \cite{Nesvornyetal2017}  is significantly wider than the observed one.

In this work, we revisit this problem and study the influence of Planet-9 on the formation and evolution of Ecliptic comets and trans-Neptunian objects. We model the formation and dynamical evolution of cometary reservoirs during Neptune's planetesimal driven migration phase. One of the main improvements of our work compared to that of \citet{Nesvornyetal2017} is that we perform a simulation considering the newly updated mass and orbital parameters of Planet-9. Simulations by \citet{Nesvornyetal2017} considered a hypothetical Planet-9 with masses of 10, 15, 20, and 30~$M_{\oplus}$. As discussed before, this proved to be largely inconsistent with ECs. Utilizing a series of numerical simulations, \citet{BB2019} demonstrated that a hypothetical Planet-9 with masses between 5 and 10~$M_{\oplus}$ generally leads to an apsidally clustered population of distant TNOs with a degree of orbital confinement more favored by observations, compared to higher-mass Planet-9 cases. By reducing the Planet-9 mass, we aim at testing if this revised mass turns out to be also more consistent with the observed population of ECs. We do not aim to infer or confirm the existence of Planet 9 using ECs; rather, this paper tests whether the orbital distribution of the EC pollution remains problematic within the context of a solar system that includes Planet 9, considering the updated orbital parameters. Our simulations also provide a self-consistent framework to understand the dynamical configuration of distant Kuiper belt objects under the influence of Planet-9.

In our simulation, we assume a Planet-9 with a mass of $m_9 \sim 7.5 M_{\oplus}$, evolving on a moderately inclined orbit with $i_9 \sim 20$ deg, with semi-major axis $a_9 \sim 600$ au, and orbital eccentricity $e_9 \sim 0.3$. Our selected orbital configuration is broadly consistent with the range of values constrained by \cite{2021ApJ...910L..20B}. Our simulation shows that a  less massive Planet-9, compared to those masses assumed for Planet-9 in \cite{Nesvornyetal2017}, indeed produces a SDOs population with relatively lower orbital inclinations. We compare the orbital distributions and number of active comets produced in our simulation with observations. Considering that comets remain active for N($q$) perihelion passages, we found that the orbital distribution of ECs is well reproduced in our models with $N($2.5$) \sim 100 - 1000$ in the presence of Planet-9. Our nominal model estimates a number of ECs broadly similar to what is observed. Finally, our simulation predicts that most extreme trans-Neptunian objects with $250 < a < 1000$~au and $q > 40$~au are apsidally aligned with Planet-9. Only objects with orbital inclinations lower than $\sim$10-20° (low-inclination objects currently dominate the observed population of trans-Neptunian objects) actually show apsidal anti-alignment with Planet-9, as usually envisioned \citep[e.g.][]{2021ApJ...910L..20B}.

\section{Methods}\label{Sec:Methods}

\subsection{Early evolution of Solar System}
\label{early}

In our simulation, we model the formation and evolution of the trans-Neptunian object population and Oort cloud  as envisioned in the most successful realizations of the so-called \textit{Nice model} \citep{NesvornyMorby2012}. Our simulations mimic the giant planets orbital radial migration during their gravitational interaction with a massive primordial planetesimal disk ($M_{disk} = 15-20 M_{\oplus}$) placed beyond Neptune's orbit \citep{Nesvorny2012, Nesvorny2016a,N2017c}. Our primordial planetesimal disk radially extends from about 22 to 30 au and all the particles in the planetesimal disk are considered test particles.

We integrate the orbits of the four giant planets (Jupiter-Neptune), the hypothetical Planet-9, and planetesimal population in the outer primordial planetesimal disk.  Following \citet{N2017c}, Jupiter and Saturn are placed initially in their current orbits, while Uranus and Neptune are initially placed inside their current orbits and then forced to migrate outward.  In this study, we set the initial orbit of Neptune as $a_{N,0} = 24$ au, $e_{N,0} = 0$, and $i_{N,0} = 0$ \citep{Nesvorny2016a}. Uranus starts with $a_{U,0} = 16$ au, $e_{U,0} = 0$, and $i_{U,0} = 0$.

Our simulation is performed using the IAS15 integrator, which is part of the REBOUND code \citep{Rein2012, ReinSpiegel2015, ReinTamayo2015}. The original code was slightly modified to include artificial forces that mimic the radial migration and damping of planetary orbits due to the interaction with the planetesimal disk. These artificial forces are parametrized by the exponential e-folding timescales $\tau_a$, $\tau_e$, and $\tau_i$, which control the rates of radial migration, inclination and eccentricity damping \citep{Nesvorny2016a}. We set $\tau_a = \tau_e = \tau_i$, as suggested by \citet{Nesvorny2012}.

Our numerical integrations are divided into two stages using different migration/damping timescales, $\tau_1$ and $\tau_2$. We set $\tau_1 = 10$ Myr and $\tau_2 = 30$ Myr, as also suggested by \citet{Nesvorny2012}.
The first stage ends when Neptune reaches $a_{N,1} = 27.7$ au. In order to approximate the effect of planetary encounters during the dynamical instability, we applied a discontinuous change to Neptune's semimajor axis ($\Delta_{a_N}$)
and eccentricity ($\Delta_{e_N}$). We used $\Delta_{a_N} = 0.5$ au and $\Delta_{e_N} = 0.1$,  based on the findings of \citet{Nesvorny2012, Nesvorny2015a, Nesvorny2015b}.
We assumed that the migration and damping timescale of Uranus are the same as those of Neptune. Since the orbit of Uranus is not significantly affected by  planetary encounters in \citet{Nesvorny2012}, we used $\Delta_{a_U} = 0$ and $\Delta_{e_U} = 0.05$.

As Neptune scatters and has close-encounters with  nearby planetesimals (planetesimal scattering) from the disk its radial migration tend to be grainy rather than ``smooth'' \citep{Nesvorny2012, Nesvorny2015a, Nesvorny2015b, Nesvorny2016a}. Grainy migration is important for explaining some specific populations of objects in the Kuiter belt, but it is not crucial for the formation of cometary reservoirs \citep{Nesvorny2016a, N2017c}. Yet, for completeness, our simulations include the effects of grainy migration following the prescription provided by \citet{Nesvorny2016a}. Our phase-1 and 2 simulations were conducted over a timescale of 500 Myrs. Subsequently, these simulations were extended up to 4.5 Gyr. Migration and damping forces were not considered after the initial 500 Myrs.

\subsection{Planet-9 parameters}

In order to develop a model that accounts for Planet 9, it is essential to understand its formation process and the mechanisms that placed it on its current orbit. The notion that Planet-9 formed in its current location faces challenges due to its slow orbital period of around 10,000 years (500 AU distance), making it difficult to grow within the typical lifetime of a protoplanetary disk \citep{kenyonbromley16}. It is more likely that Planet-9 originated closer to the known giant planets, possibly between 5 to 30 AU, before being ejected outward \citep{Izidoroetal2015}. Some studies have explored the possibility that additional ice giants were ejected during the solar system dynamical instability \citep{Nesvorny2011,Nesvorny2012}. It remains unclear whether Planet 9 is directly connected to the hypothesis of a fifth giant planet \citep{Nesvorny2011,Nesvorny2012}. Particularly if the planetary instability occurred relatively late as, for example, at $>$10 Myr after the solar system's formation, it becomes challenging to envision a mechanism stabilizing a fifth planet on a wide orbit. Instead, it appears more feasible that Planet-9 settled into its wide orbit well before the era of planetary instability \citep{Izidoroetal2015,10.1093/mnras/stad3017}. This suggests that the origin of the trans-Neptunian populations, including distant cometary reservoirs, came after the events leading to Planet-9's wide orbit. Under this premise, we assume that Planet-9 is already in place during the  migration/instability phases of Neptune and Uranus.

Due to its computation cost, we studied a single set of parameters for Planet-9.  In our simulation, Planet-9 is assumed to have a  mass of $m_9 = 7.5 M_{\oplus}$, orbital inclination of $i_9 = 20$ degrees,  semi-major axis of $a_9 \sim 600$, and orbital eccentricity of $e_9 \sim 0.3$. As discussed before, the selected set of orbital parameters for Planet-9 used in our simulation is consistent with those suggested by \citet{BB2019}.

\subsection{Primordial Planetesimal Disk}

In our simulation, the primordial disk of planetesimals initially extends from $\sim 22$ au to $r_{\text{edge}} = 30$ au. Our simulation is high resolution (starting with $10^6$ test particles). Planetesimals are treated as massless particles. The initial eccentricities and initial inclinations of the disk particles followed a Rayleigh distribution with $\sigma_e = 0.1$ and $\sigma_i = 0.05$, where $\sigma$ represents the usual scale parameter of  Rayleigh distributions.

\subsection{Effect of the Galactic tidal forces and Stellar Encounters}
\label{external}

Our simulation also account for the effects of Galactic tides and passing stars.  We implemented the effects of Galactic tidal forces and stellar encounters in the REBOUND code using the same approach as described in \citet{N2017c}.
We assume that the Galaxy is axisymmetric, and the Sun follows a circular orbit with a radius of $R_o$ within the plane of our Galaxy.
The angular velocity of the Sun is denoted as $\Omega_o$ relative to the center of the Galaxy.
The tidal acceleration from the Galaxy is given by \citep{1986Icar...65...13H,1999Icar..137...84W,2001AJ....121.2253L}:

\begin{equation}
\vec{a}_{tide} = \Omega_o^{2} \left[ (1 - 2 \delta) x \hat{e}_x - y \hat{e}_y - \left( \frac{4 G \pi \rho_{o}}{\Omega_o^{2}}   - 2 \delta \right) z \hat{e}_z \right],
\end{equation}
 where $\delta = - \frac{(A+B)}{(A-B)}$, and A and B are the Oort constants. G is the gravitational constant, and $\rho_{o}$ is the mass density in the vicinity of the Sun.
The coordinate system ($\hat{e}_x, \hat{e}_y, \hat{e}_z$) is oriented to maintain the reference frame of the integration, with the Z-axis pointed towards the initial angular momentum of the planets.
The effects of the Galactic tidal forces were modelled using an density of $\rho_{o}$ = 0.1 $ \textup{M}_\odot pc^{-3}$, and Oort constants set to  A = 14.82 $kms^{-1} kpc^{-1}$ and
B = -12.37 $km^{-1} kpc^{-1}$, and $\Omega_o = 2.8 \times 10^{-8} yr^{-1}$ \citep{N2017c}.

The effect of stellar encounters was implemented in the REBOUND code by adding a star at the beginning of each close encounter and removing the star once the encounter ended.
Stars were removed when their heliocentric distances reached 206,000 astronomical units or 1 parsec. To generate stellar encounters, we used the model proposed by \citet{1986Icar...65...13H}.
The stars are of various types in terms of mass and velocity, including main sequence stars and white dwarfs. We utilized approximately 47,000 stellar encounters for our simulations.
We do not consider the early stages when the Sun likely interacted with other stars within its natal cluster \citep{2010ARA&A..48...47A}.
However, it is worth noting that during the early stages, planetesimals in the size range of comets were significantly influenced by
aerodynamic gas drag and may not contribute to the formation of cometary reservoirs \citep{N2017c}. Therefore, our scenario corresponds to the case where the disk lifetime is equal or longer than the cluster lifetime.

\subsection{Observed short period comets}
\label{observations}

In order to compare the results of our model with observations, we only account for active comets \footnote{Inactive/dormant comets are not considered because only a few are known.} with a known absolute magnitude ($H_t$) that contain both a nucleus and coma.
This approach is invoked in order to reduce the impact of observational biases in our analysis. Additionally, we removed from the observed sample paired bodies, such as fragments, to ensure that only one data point remained for each parent comet.

Our observed sample of comets was obtained from the JPL Small-Body Database Search Engine in January 2023. In order to  filter the sample for ECs, we follow \citet{1997Icar..127...13L}, where short-period comets were classified into two categories based on their respective Tisserand parameter \citep{N2017c}: ecliptic comets (ECs) with $2 < T_J < 3$ and nearly isotropic comets (NICs) with $T_J < 2$.  We restrict our analysis to cometary orbits with orbital periods less than 20 years ($P < 20$ yrs) and Tisserand parameter values ranging from 2 to 3 ($2 < T_J < 3$).


\begin{figure*}
 \includegraphics[scale = 0.1]{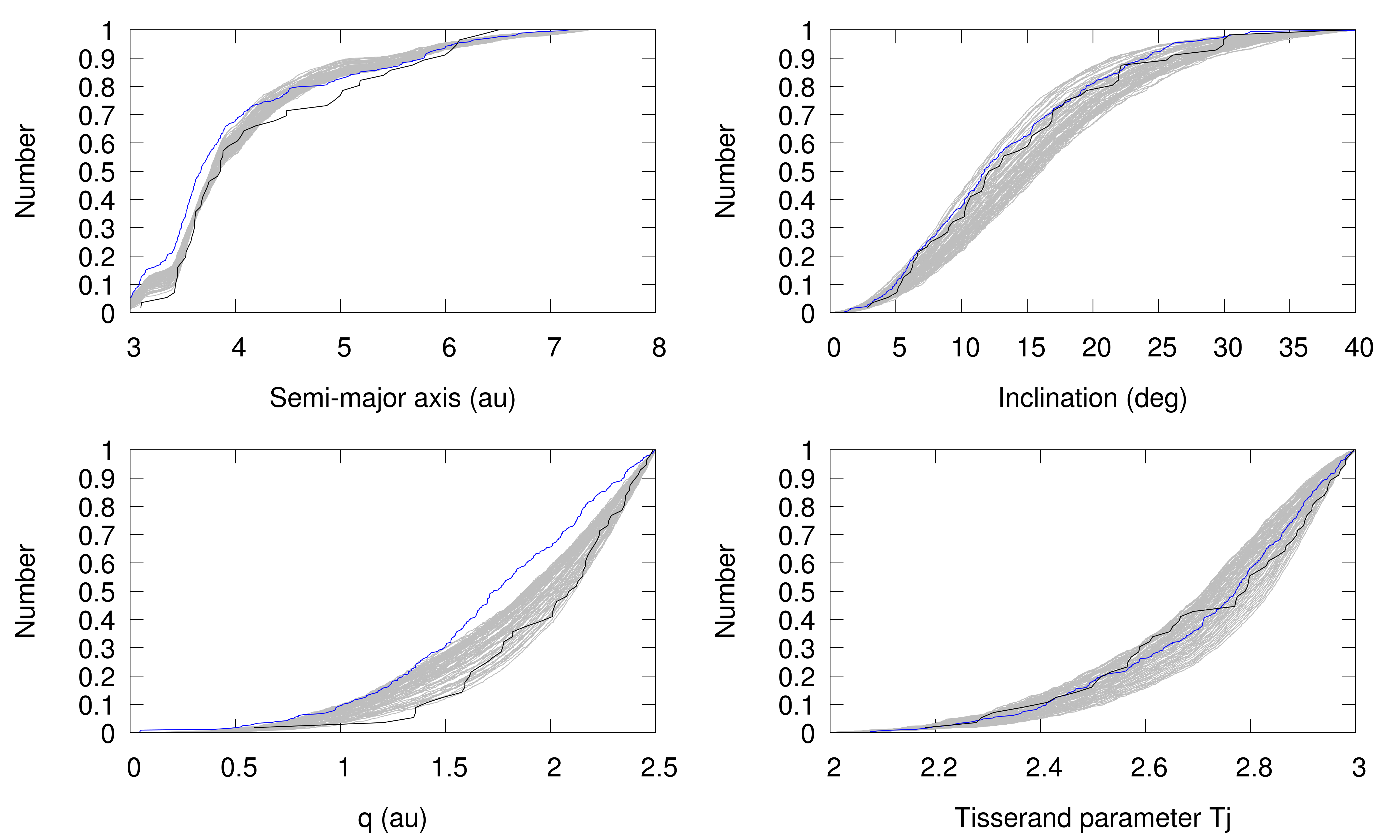}
\caption{Normalized cumulative orbital distributions of  ECs satisfying the criteria $P < 20$ yrs, $2 < T_J < 3$, and $q < 2.5$ au. The black curves show the distributions of observed ECs with total magnitude $H_t < 10$, consisting of a total of 57 objects. Blue curves show the distributions of all observed ECs, which consists of a total of 84 objects. Gray curves represent the orbital distributions of comets produced in our simulations considering that ECs comets remain active for a number of perihelion passage ($N(q<2.5~{\rm au}$)) ranging from 150 to 1000. Each gray curve represents one specific value of physical time $N(q<2.5~{\rm au}$).}
\label{fig:observational}
\end{figure*}

The normalized cumulative distribution of the orbital elements of known ECs satisfying the criteria $P < 20$ yrs, $2 < T_J < 3$, and $q < 2.5$ au are shown in Figure \ref{fig:observational}. Black curves represent the orbital distributions with a cut in $H_t < 10$ (a total of 57 ECs), while the blue curves represent distributions without any cut in $H_t$ (a total of 84 ECs). It is worth noticing that, as emphasized in \citet{N2017c}, the orbital distribution of ECs is reasonably well-defined within this parameter range. The distributions of semimajor axis, inclination, and Tisserand parameter are nearly independent of any cutoff in total absolute magnitude.


In the next section, we explain how we will study and analyze the formation and evolution of cometary reservoirs in our simulation.

\subsection{Production of Cometary orbits}
\label{production}

To study the evolution of  cometary reservoirs in our simulation, we follow the approach of \citet{N2017c}. During the final integration phase of our simulation, which spans  from 3.5 Gy to 4.5 Gy, we utilize a specific code designed to track, analyze, and record cometary orbits. Its algorithm works as the following.

We continuously monitor the heliocentric distance, r, for each planetesimal at every time step in our simulation. In order to improve the statistical significance of orbits below Saturn's orbit, when an object reaches $r < 9$ au for the first time, we clone it. Each object satisfying this condition is cloned 100 times, resulting  in 100 new objects. Cloned objects are created by applying small random perturbations to the velocity vector of the original object, as outlined in \citet{BrasserandMorbidelli2013} and further discussed in \citet{sousaribeiro2022}.

We record the orbital elements of comets in our simulations at intervals of 100  years. The orbital evolution of comets is also recorded if their semimajor axis ($a$) is smaller than 35 au, corresponding to a period ($P$) of approximately 200 years, and the perihelion distance ($q$) is less than 5.2 au. We  rotate the orbits of these objects to the reference plane defined by the instantaneous angular momentum vector of the four outer planets (for more detailed information, please refer to \citet{N2017c}).
In our simulations we neglect non-gravitational forces acting on comets.

In order to compare the orbits of comets produced in our simulations with observations, we treat the typical lifetime of active comets  as free parameter. A comet becomes extinct when it has expelled virtually all its volatile ice and can not effectively sustain a tail and coma. The detailed discussion on this issue can be found in \citet{1997Icar..127...13L} and \citet{N2017c}. In this study, a comet is considered extinct after a given number of perihelion passages with $q<2.5$~au, represented by $N_p(2.5~au)$.

\section{Results }\label{Sec:Results}

\subsection{Dynamical formation of cometary reservoirs}
\label{planetesimal_evol}

\begin{figure*}
 \includegraphics[scale=0.09]{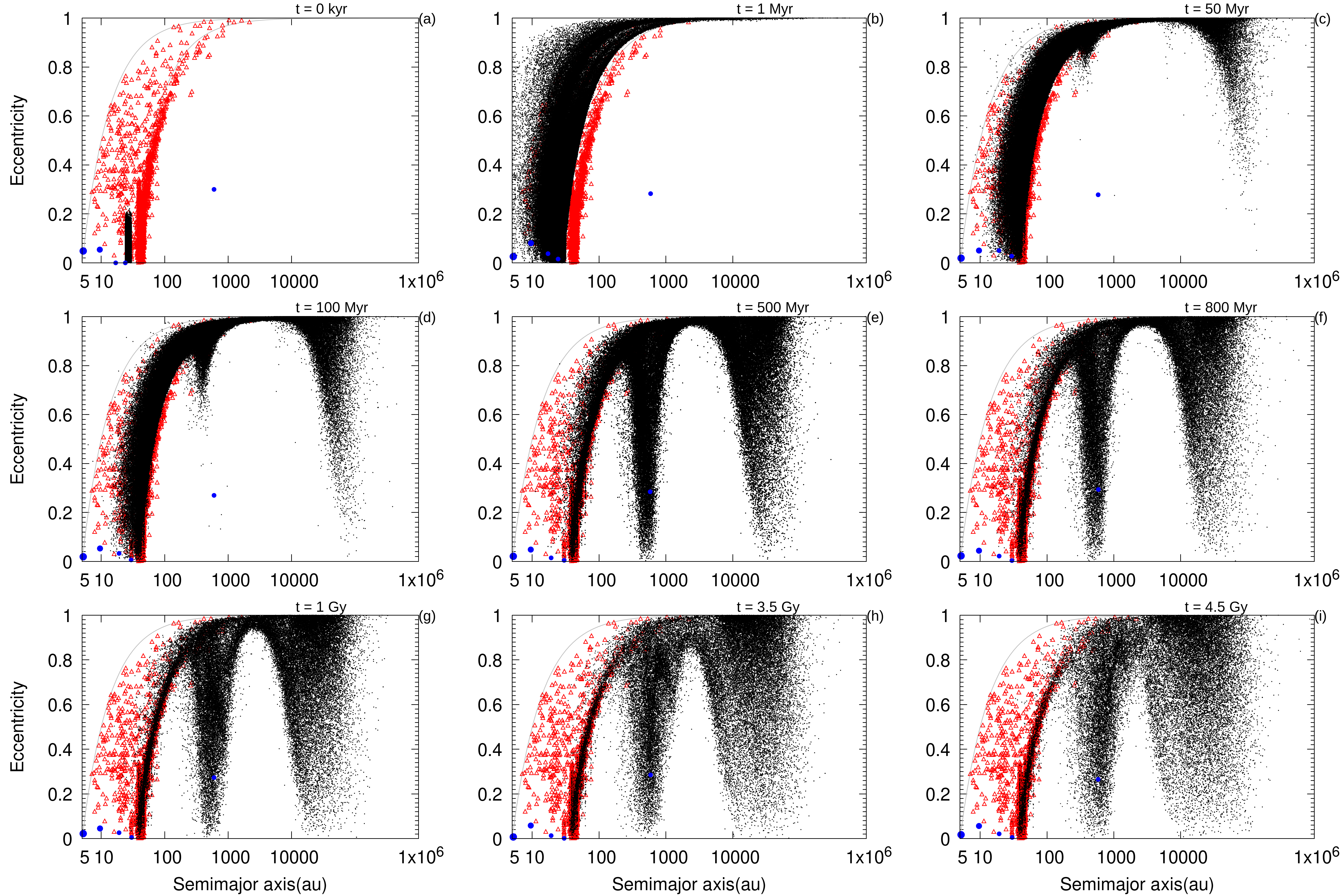}
\caption{Snapshots showing the formation and evolution of the scattered disk objects, Planet-9 cloud, and Oort cloud. The x-axis show semi-major axis and the y-axis show orbital eccentricity. Simulations were performed considering the effects of  Galactic tides, with  $\rho_{o}$ = 0.1 $ \textup{M}_\odot pc^{-3}$, and the effects of stellar encounters. Planets are shown as blue dots. The red triangles represent observed Kuiper belt and Centaurs objects.} \label{fig:case_a_e}
\end{figure*}

\begin{figure*}
 \includegraphics[scale=0.09]{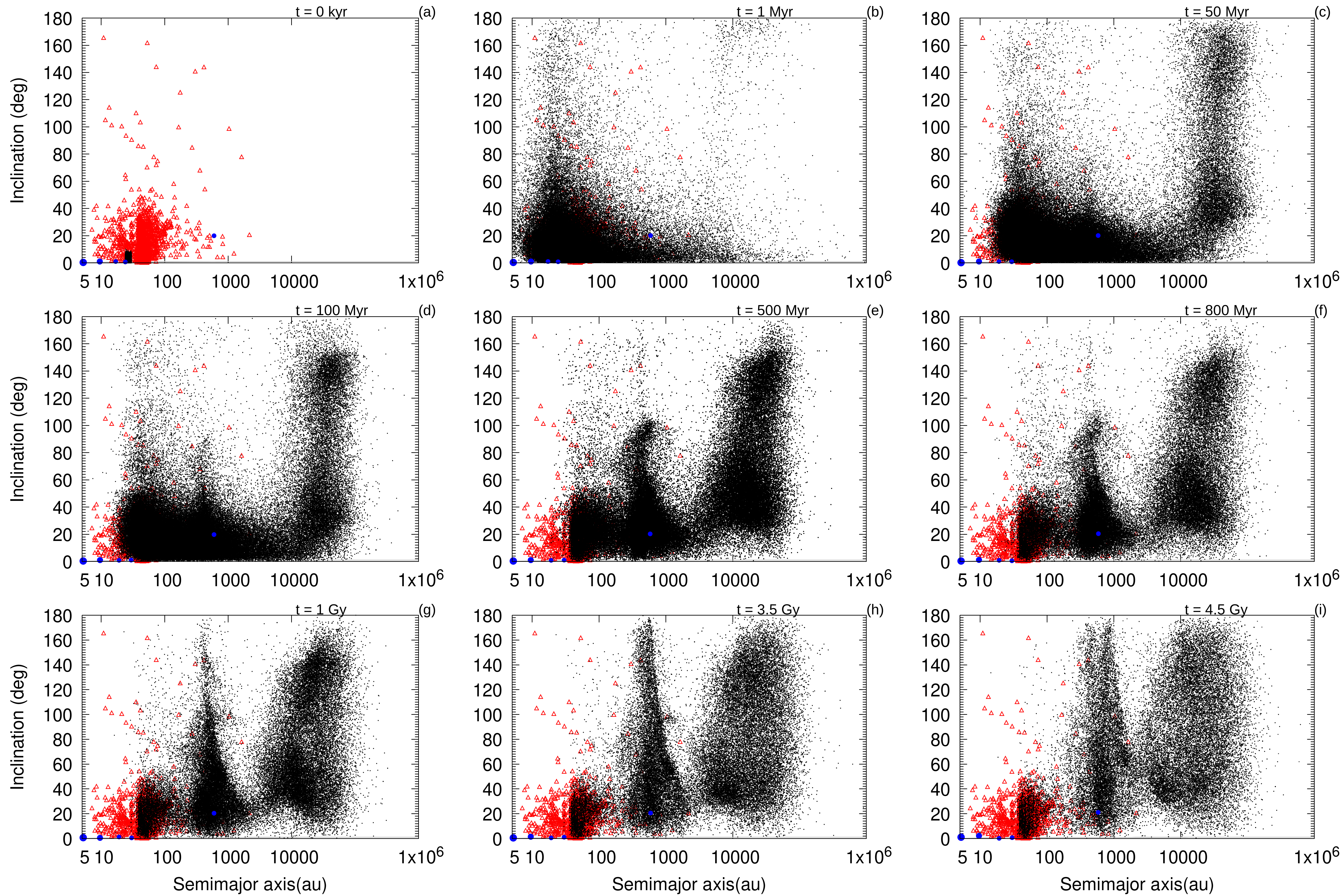}
\caption{The same as Figure \ref{fig:case_a_e} but for orbital inclination.}
\label{fig:case_a_i}
\end{figure*}

Figures \ref{fig:case_a_e} and \ref{fig:case_a_i} show the orbital structure of the trans-Neptunian region generated in our simulation at 0, 1, 50, 100, 500, 800 Myr, 1 Gy, 3.5 Gy, and 4.5 Gy.
As discussed in \citet{2016AJ....152..133K, Nesvorny2016a, N2017c}, the first structure to form in the trans-Neptunian region is the scattered disk, which we defined before as the region where $50 < a < 1000$ au.   The \textbf{SDOs} originates when bodies interact with Neptune
through close encounters at approximately $q \sim 30$ au, resulting in their scattering to very large heliocentric distances.

Table \ref{tab:distribuicao_sdo} shows the number of objects across different semimajor axis intervals based on their location and orbital inclination at 4.5 Gy. $N\_Disk$ accounts for objects in the disk component ($i < 30^\circ$) and $N\_Halo$ accounts for those in the halo component ($i > 30^\circ$). We define four different regions as: Inner SDO ($50 < a < 200$ au), P9 Cloud / Outer SDOs ($200 < a < 1000$ au), Inner OC ($1000 < a < 20000$ au), and Outer OC ($a > 20000$ au). The bottom line of Table \ref{tab:distribuicao_sdo} shows objects within these same regions but with $q < 38$~au and $q > 38$~au.

\begin{table*}
\centering
\begin{tabular}{|c|c|c|c|c|}
\hline
\textbf{Category} & \textbf{Inner SDO} & \textbf{P9 Cloud / Outer SDOs} & \textbf{Inner OC} & \textbf{Outer OC} \\
                 & \textbf{($50 < a < 200$ au)} & \textbf{($200 < a < 1000$ au)} & \textbf{($1000 < a < 20000$ au)} & \textbf{($a > 20000$ au)} \\
\hline
\hline

\textbf{$N\_Disk$ ($i < 30^\circ$)}             & 3226 & 5726 &  3844 & 1753 \\
\textbf{$N\_Halo$ ($i > 30^\circ$)}             & 1980 & 21999 & 39310 & 20589 \\
\hline
\hskip4in
                 & \textbf{($50 < a < 200$ au} & \textbf{($200 < a < 1000$ au } & \textbf{($1000 < a < 20000$ au} & \textbf{($a > 20000$ au} \\
                 & \textbf{ and $q < 38$ au)} & \textbf{and $q < 38$ au)} & \textbf{and $q > 38$ au)} & \textbf{ $q > 38$ au)} \\
\hline
\textbf{All}             & 3285 & 1207 &  42136 & 22295 \\
\hline
\end{tabular}
\caption{Number of objects in different regions at 4.5 Gyr (end of our simulations). \textbf{$N\_Disk$} refers to the number of objects in the disk component with inclination of less than 30$^\circ$ ($i < 30^\circ$), while \textbf{$N\_Halo$} corresponds to the halo component, objects with inclination greater than 30$^\circ$ ($i > 30^\circ$). The regions are defined as follows: \textbf{Inner SDO} refers to scattered disk objects with $50 < a < 200$ au, \textbf{P9 Cloud / Outer SDOs} covers scattered disk objects in the range $200 < a < 1000$ au, \textbf{Inner OC} refers to objects in the outer Oort cloud with $1000 < a < 20000$ au, and \textbf{Outer OC} corresponds to the outermost objects in the Oort cloud with $a > 20000$ au. The bottom row shows the number of objects with orbital parameters defined at the top of the table, categorized by $q < 38$~au or $q > 38$~au.}

\label{tab:distribuicao_sdo}
\end{table*}


The formation of the inner portion of the scattered disk (hereafter \textbf{inner SDOs}, $50 < a < 200$ au) is dependent on Neptune's migration \citep{2016AJ....152..133K, Nesvorny2016a}.
We identified approximately 5206 bodies at $t = 4.5$ Gy that are located within the inner scattered disk.
This total represents a fraction of about $5 \times 10^{-3}$ of the initial  planetesimal population ($10^6$). Envisioning a primordial planetesimal disk of about $20 M_{\oplus}$, it yields approximately 0.10 $M_{\oplus}$. If we included only objects with $q < 38$ au and $50 < a < 200$ au, the total number of objects is 3285, with a total mass of 0.06 $M_{\oplus}$.
The simulations of \citet{N2017c} performed with and without  Planet-9 resulted in a total of 0.06 and 0.05 $M_{\oplus}$ in planetesimals within the \textbf{inner SDOs}, respectively. We recall that in \citet{N2017c}, Planet-9  mass was set to $m_9 = 15.0 M_{\oplus}$. Therefore, overall, the presence of Planet-9 only slightly affects the total mass (and number) of objects within the \textbf{inner SDOs}.

In our simulation, $\sim 1500$ bodies are in stable orbits within the classical Kuiper Belt (with $a<50$ au), encompassing both the hot and resonant populations.
This corresponds to a fraction of approximately 0.03 $M_{\oplus}$ in terms of mass. This mass is strikingly consistent with the value derived from ephemerides (as reported by \citet{2018CeMDA.130...57P}),
estimated to be approximately $2 \times 10^{-2} M_{\oplus}$. Interestingly, this value closely aligns with the results of simulations conducted without Planet-9 by \citet{N2017c}. This indicates that the presence of Planet-9 does not impact the orbital
structure or the number of objects within the classical Kuiper Belt.

In our simulation, we observed that starting from 50 Myr (see in Fig. \ref{fig:case_a_e} (c)), the orbital structure of the trans-Neptunian region undergoes changes due to the gravitational influence of Planet-9.
Planet-9's gravitational influence causes orbital changes in SDOs. Through secular dynamics and mechanisms like Kozai-Lidov resonance, Planet-9 can increase an SDO object perihelion distance effectively detaching it from Neptune's influence. This results in the formation of a nearly isotropic cloud of bodies whose semimajor axes are centered around the semimajor axis of Planet-9, known as the \textbf{P9 cloud} (as also described in \citet{N2017c}). We found approximately $ \sim 2.7 \times 10^{4}$ objects located within the \textbf{P9 cloud} (defined here as $200 < a < 1000$ au) at 4.5 Gy. This accounts for a fraction of 2.7\% of the initial number of planetesimals. In terms of mass, this corresponds to approximately 0.55 $M_{\oplus}$ for a planetesimal disk with a total mass of $20 M_{\oplus}$.
When comparing our results with the simulations conducted by \citet{N2017c}, we observed a population that is 1.47 times larger than what was found in their study. This discrepancy can be attributed to the presence of a less massive and eccentric Planet-9 in our simulation. Such a configuration allows for the maintenance of a significant concentration of objects with eccentricities smaller than 0.3, in contrast to the findings reported in \citet{N2017c}.

We define the region with semimajor axes ranging from 1000 to 20,000 au as the \textbf{inner Oort cloud}, while the \textbf{outer Oort cloud} encompasses the region with semimajor axes exceeding 20,000 au.
The \textbf{outer Oort cloud} starts to form within the first 10 million years, primarily through close encounters with Saturn, Uranus, and Neptune. In contrast,
Jupiter, which primarily acts as an ejector of planetesimals in the Solar System \citep{2004come.book..153D, Dones_2015}, does not contribute significantly to the formation of the \textbf{outer Oort cloud}. In our simulations, most bodies that ended up in the inner Oort Cloud were scattered to $a > 1000 \, \text{AU}$ by Neptune or Uranus. Planet-9 also appears to facilitate the decoupling of scattered objects from Neptune, further increasing the populations of bodies in the Planet-9 cloud and the inner Oort Cloud.

In our simulation, we observe that approximately $6.5 \times 10^{4}$ objects populate the Oort cloud at 4.5 Gy. This accounts for roughly 6.5\%
of the original number of planetesimals in the simulation ($\sim 1.1 M_{\oplus}$). This finding aligns with the results reported in \citet{N2017c} and \citet{BrasserandMorbidelli2013}, and is slightly higher than estimates derived from
dynamical models by \citet{2004come.book..153D} and \citet{Dones_2015}, which suggest a range of 3-5\%.
The majority of objects are found in the inner Oort cloud, accounting for 65\% of the total, while the remaining 35\% reside in the outer Oort cloud.

The ratio between the number of objects in the Oort cloud (defined as $a > 1000$~AU and $q > 38$~AU) and those in the scattered disk (for better comparison with previous studies here it is defined as objects with $50 < a < 200$~AU and $q < 38$~AU) in our simulations, including Planet-9, is approximately 20. The specific counts for each region are provided in Table \ref{tab:distribuicao_sdo}. In the absence of Planet-9, previous studies by \citet{BrasserandMorbidelli2013} and \citet{N2017c} found this ratio to range between $\sim$12 and $\sim$20. Our result aligns well with the observationally favorable value of approximately 20, as reported by \citet{Krolikowskaetal10}.


\subsection{Orbital distribution of ecliptic comets}
\label{ecliptic}

In this section, we compare the orbital distribution of ECs produced in  our simulation with the orbital distribution of real ECs comets (see Sections \ref{observations} and \ref{production}) using a statistical test. In our analysis, we only considered comets with orbital distributions that satisfy the criteria of $P < 20$ yrs, $2 < T_J < 3$, and $q < 2.5$ au.

\begin{figure*}
    \includegraphics[scale=0.055]{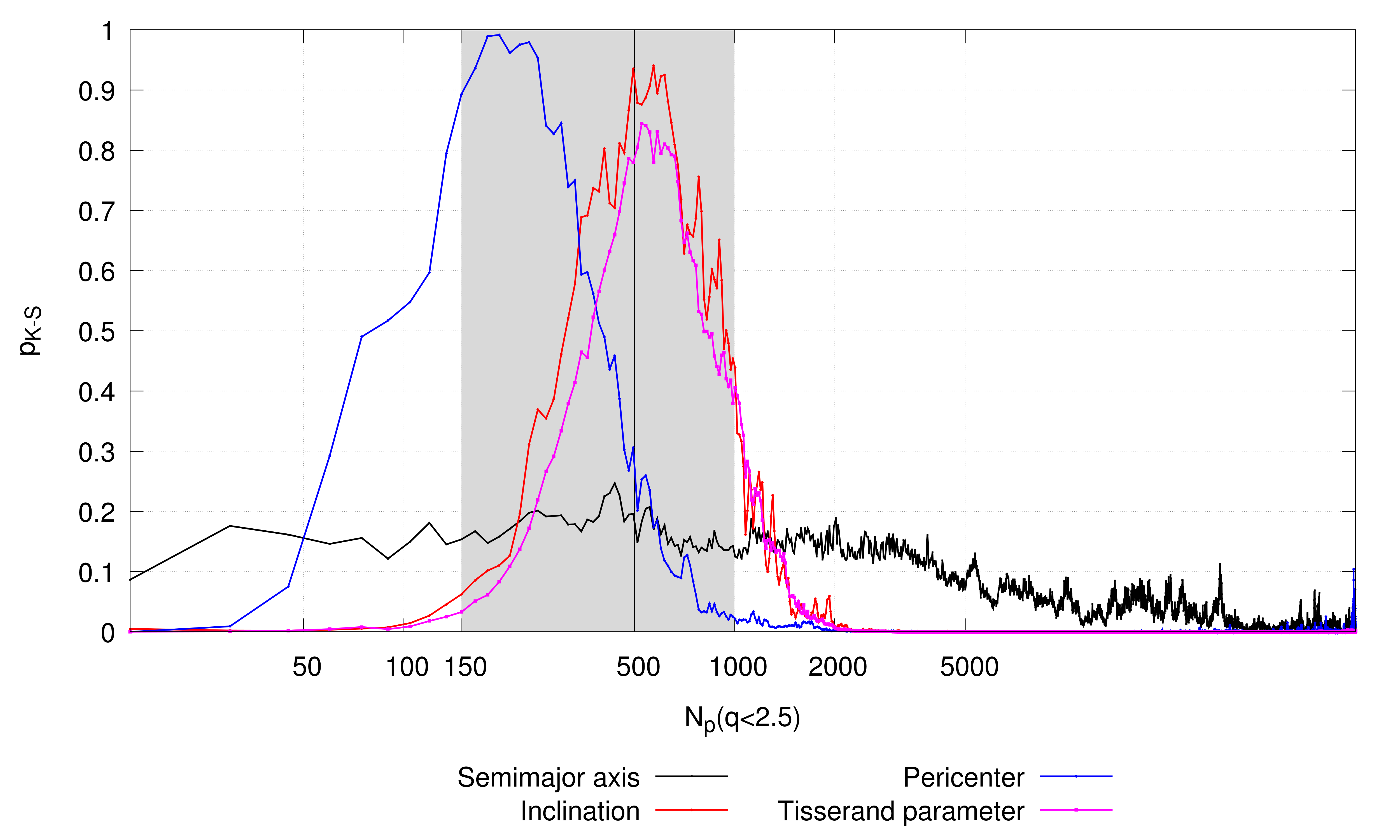}
    \includegraphics[scale=0.055]{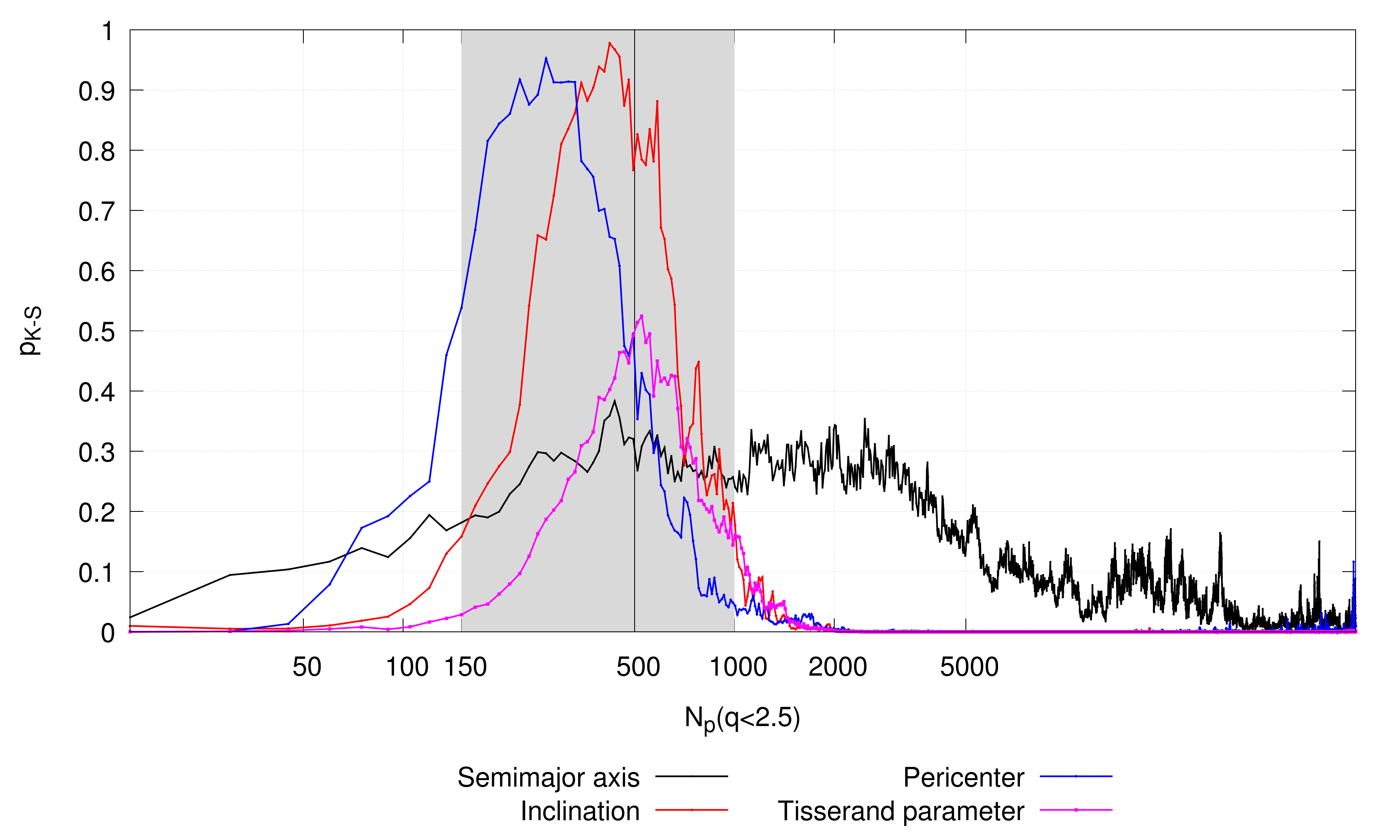}
   \caption{Statistical KS-tests comparing the distribution of ECs in our model and observations. $p_{K-S}$-values are reported as a function of the number of perihelion passages with $q < 2.5$ au which a comet is assumed to become extinct. The $p_{K-S}$ values for  semimajor axis, pericenter, inclination, and Tisserand parameter are given in black, blue, red, and magenta colors, respectively. Following the definition of ECs, in out analysis, we only considered comets with orbital distributions that satisfy the following criteria: $P < 20$ yrs, $2 < T_J < 3$, and $q < 2.5$ au. The left panel show the result of our KS-tests for an observed distribution with a cut in $H_t < 10$ (a total of 57 ECs), while the right panel shows results without any cut in $H_t$ (a total of 84 ECs). In both panels, the gray region corresponds to the number of perihelion passages where our model provides a very good match to the orbital distribution of real comets, with $p_{K-S}$-values typically much larger than 0.1~.  The best fit considering all four orbital distributions corresponds to approximately 500 perihelion passages with $q < 2.5$ au.}
    \label{fig:ks_tests}
\end{figure*}

Figure \ref{fig:ks_tests} shows the outcome of our KS-tests comparing the real and simulated ECs. The grey region in Figure \ref{fig:ks_tests}  shows that the p-values of our KS-tests --  for different orbital elements -- broadly peak with the assumption of between 150 and 1000 perihelion passages with $q<2.5$~au for a comet to become extinct. The orbital distribution of ECs in our simulations, considering this range of perihelion passages for extinction (150 to 1000), are presented in Figure \ref{fig:observational} as grey lines.

Our KS-tests show that the orbital distributions of ECs produced in our simulations are broadly consistent with observations. This result is different from  the results of \citet{N2017c}, where simulations including a more massive Planet-9  produced ECs with broader inclination distributions than the observed one.



The inner SDOs (Scattered Disk Objects), defined as objects with semi-major axis between $50 < a < 200$ au are the primary source of ECs \citep{N2017c}. The presence of  Planet-9 leads to interactions with the inner SDOs, causing their inclinations to become dynamically more excited than when Planet-9 is not present \citep{N2017c}. In our simulation, Planet-9 is present but is relatively less massive. Consequently, it is not capable of over-exciting the inner SDOs as a Planet-9 with $m_9 \gtrsim 10.0 M_{\oplus}$, as invoked by \citet{N2017c}, does.

\begin{figure}
\centering
    \includegraphics[scale=0.035]{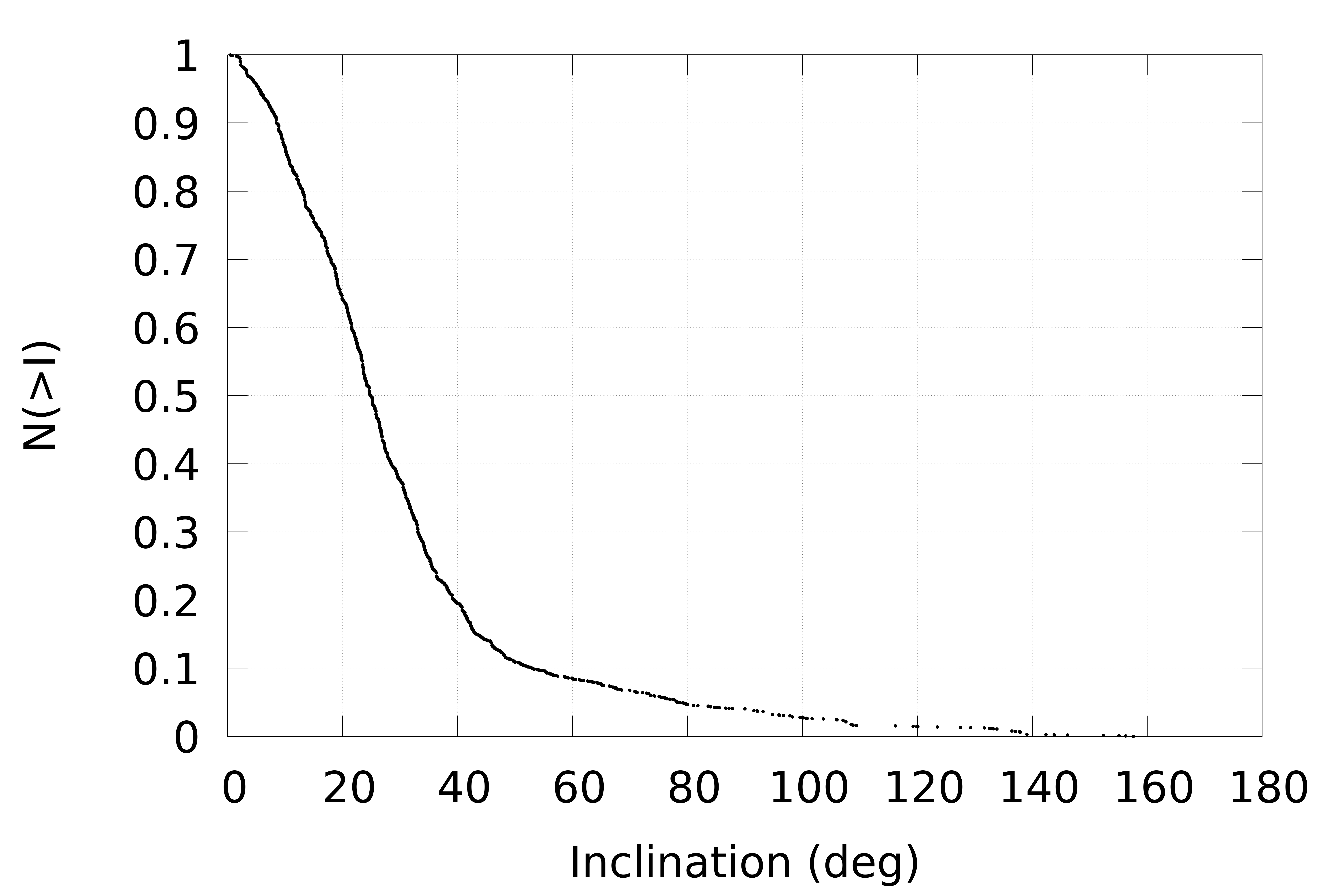}
       \caption{Normalized cumulative inclination distribution of the inner SDOs (with semi-major axes between 50 and 200 au) at 4.5 Gyr.}
    \label{fig:number_inclination}
\end{figure}

Figure \ref{fig:number_inclination} shows the cumulative distribution of inclinations of the inner SDOs in our simulation at 4.5 Gy. Our simulation shows that approximately 45\% of the objects in the inner SDOs have inclinations between 20 and 40 degrees, about 40\% have inclinations greater than 30 degrees,  and  20\% have inclinations greater than 40 degrees. For comparison, the inclination distributions reported in \citet{N2017c} with a more massive Planet-9 have 60\% of the inner SDOS objects with inclinations greater than 30 degrees, and 50\% of these objects have inclinations greater than 40 degrees. This clearly shows that the inclination distribution of the inner SDOs in our simulation with a  less massive Planet-9 is relatively flatter compared to the distributions observed in the simulations conducted by \citet{N2017c} with a more massive Planet-9.

\subsection{The expected number of ECs}
\label{number}

We will now estimate the number ECs  that would be produced in our model and compare our results with observations. Previous studies have estimated the number of ecliptic comets using numerical simulations neglecting the effects of Planet-9. \cite{1997Icar..127...13L}, for instance, estimate the number of ecliptic comets larger than $D>2$ km and $q<2.5$ au as 108, similar to the estimates of \cite{DiSisto2009} which found 117 active comets with $q<2.5$ au and $D>2$ km in their simulations. \cite{BrasserandMorbidelli2013} estimate the number of ecliptic comets with $q<2.5$ au and $D>2.3$ km to be 117$\pm50$. As pointed out in \citet{N2017c}, comets with small nuclei are probably more difficult to detect than those with $D > 10$ km and are more strongly affected by observational bias. The current observed number of ecliptic comets with total magnitude $H_T<9$ (which roughly corresponds to $D>2$ km; see \cite{BrasserandMorbidelli2013}) is about 35, a sample probably fairly incomplete. Consequently, in this work, we mainly restrict our analysis to objects with $D > 10$~km.

Observations suggest that the number of observed ECs with $D > 10$ km and $q < 2.5$ au is about  4 \citep{2004come.book..223L,2013Icar..226.1138F,N2017c}. It is also not entirely clear how complete these observations are.  For simplicity and consistency with \citet{N2017c},  we will compare the results of our simulation with observations using this number as nominal reference. Given that this comparison relies on small-number statistics, it may need to be revisited if the number of observed ECs increases in the future. For a detailed discussion on this issue, we also refer the reader to \citet{N2017c}.

In our simulation, we do not assign a  specific initial size or an initial size frequency distribution for our  primordial disk planetesimal population. In order estimate the number of ECs with $D > 10$ km we need to do so. We reconstruct the expected size frequency distribution of the primordial planetesimal disk ($a<30$~au) following the procedure of \citet{Nesvorny2016a}. Their method consists of  invoking a series of dynamical and observational constraints to infer the initial size distribution of the planetesimal population. The first constraint comes  from Neptune's migration history.

\begin{figure}
    \includegraphics[width=\columnwidth]{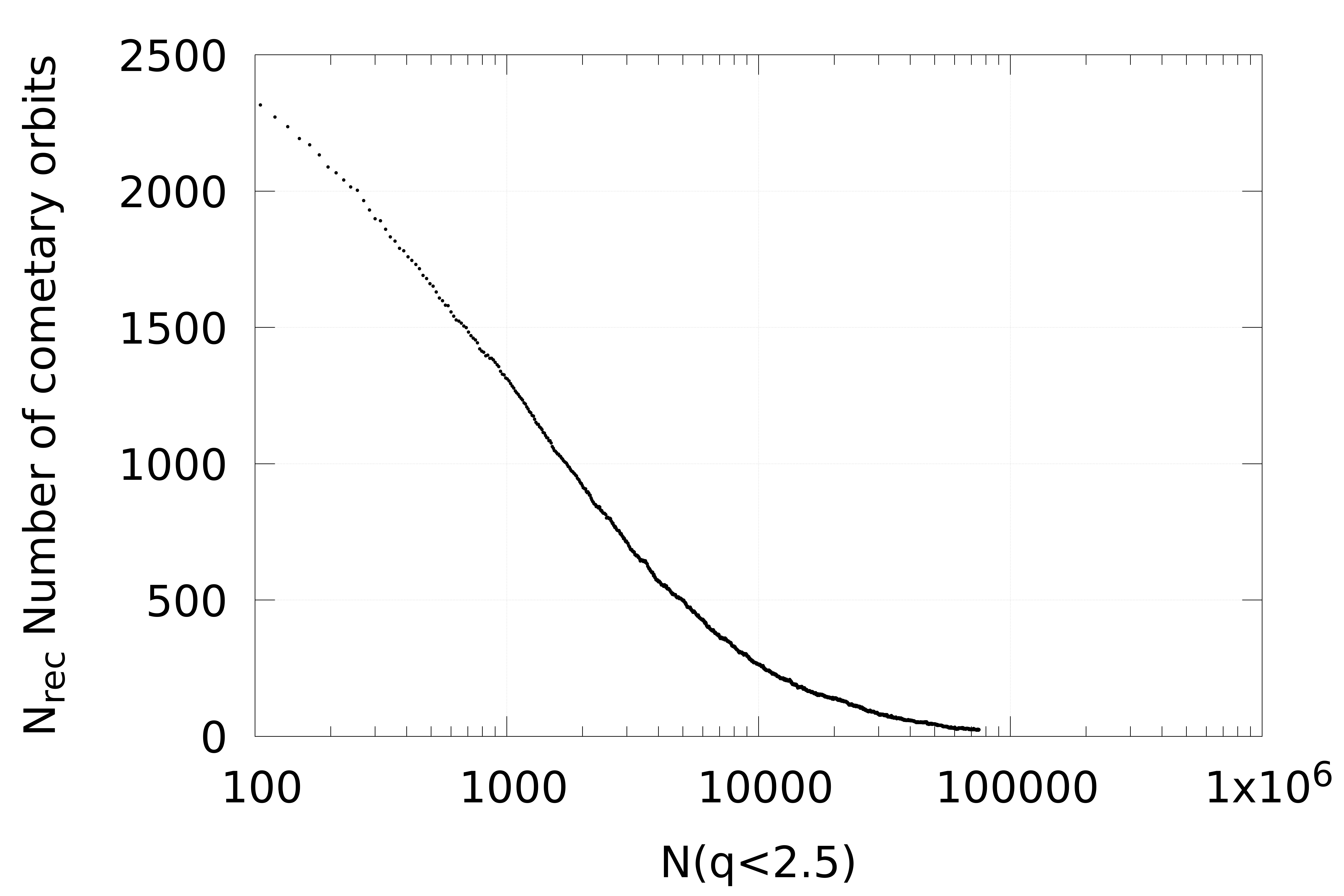}
   \caption{Number of ECs orbits recorded in our simulation as a function of the number of perihelion passages with N($q < 2.5$ au). The number of  recorded orbits decrease with the number of perihelion passages because comets are lost during the evolution of the system. When counting the number of  cometary orbits, we neglect that the fact that comets are expected to become extinct after a given number of perihelion passages.}
    \label{fig:number_ecs}
\end{figure}

In order to explain the relatively low number of objects in resonance with Neptune in the Kuiper belt today, \citet{Nesvorny2016a} argued that the planetesimal disk consisted of 1000-4000 Pluto-sized objects ($D\sim1000$~km). The gravitational interaction of Neptune with these objects makes its outward migration grainy, preventing capture and long-term stability of objects swept by mean motion resonances as it moves outwards. The second constraint comes from observations. In order to estimate the number of objects with intermediate sizes ($10 < D < 500$ km), \citet{Nesvorny2016a} used the size frequency distribution derived by \citet{Fraseretal2014} from observations of the Kuiper Belt and Jupiter Trojans. Finally, the third constraint comes from the fact that the total mass of the primordial planetesimal disk is finite. With these constraints in mind one can roughly reconstruct the the initial size frequency distribution of the primordial planetesimal disk (for additional details see \citet{Nesvorny2016a}).

\citet{Nesvorny2016a}  argue  that there were approximately $6 \times 10^9$ planetesimals in the planetesimal disk with diameters larger than 10 km. This estimate has an uncertainty of a factor of $\sim$2, primarily due to the uncertainty in the implantation probability of Jupiter Trojans and its dependence on planetary migration \citep{N2017c}. With these caveats in mind, the number of ECs larger than a given diameter $D$ can be estimated as \citep{N2017c}:
\begin{equation}
\label{eq:discummm}
N_{ECs}(D> 10 km) = N_{rec} \frac{N_{disk}(D> 10 km)}{N_{sim}} \frac{\Delta t}{\Delta T},
\end{equation}
\noindent where $N_{\text{rec}}$ is the number of recorded ECs orbits in $\Delta T$. $N_{\text{disk}}(D> 10 km)$ is the number of  planetesimals in the primordial disk with diameter larger than 10 km. $N_{\text{sim}} = 10^{8}$  is
the number of planetesimals used in our simulation including clones, $\Delta t = 100$ years is our sampling interval, and $\Delta T = 1$ Gyr is the interval used to infer the number of comets.

Figure \ref{fig:number_ecs} shows the number of recorded ECs orbits ($N_{\text{rec}}$) in our simulation as a function of the number of perihelion
passages with $q < 2.5$ au. By recalling that our simulation matches fairly well observations for $N_p(2.5)$ between  150 and 1000, the integrated total number of recorded ECs orbits in our simulation (we integrate the curve of Figure \ref{fig:number_ecs}) is approximately $3 \times 10^5$. 

Therefore, we find $N_{\text{ECs}}(D> 10 km) = 1.8-3.6$ for $D > 10$ km, $P < 20$ yrs, and $2 < T_J < 3$. This number is somewhat larger than those found by  \citet{N2017c} in simulations with and without a relatively more massive Planet-9,  where $N_{\text{ECs}}(D> 10 km) = 1-2$ and $N_{\text{ECs}}(D> 10 km) = 0.7-1$ were found, respectively (see Table \ref{tab:ecs}). This may suggests that our model, with a lower mass Planet-9, can  better reproduce the current number of observed ECs, although it is not clear how complete observations are.  By using instead our entire sample of  recorded EC orbits (about $8 \times 10^5$), we estimate a total of $N_{\text{ECs}}(D> 10 km) = 4.8-9.6$. Our simulation predicts that the relatively low mass attributed to Planet-9 leads to a significant increase in the occurrence of ECs compared to the results of  \citet{N2017c} (with a relatively more massive or without Planet-9). If we assume that the primordial planetesimal disk contained initially between $2 \times 10^{11}$ and $2 \times 10^{12}$ objects larger than 2~km, the number of ECs with $q<2.5$ au and $D>2$ km produced in our simulations with Planet-9 would be a factor of 2 to 3 higher than that estimated by previous studies \citep{1997Icar..127...13L,DiSisto2009,BrasserandMorbidelli2013} neglecting the effects of Planet-9.

\begin{table}
	\centering
	\caption{Number of ECs produced in simulations with and without Planet-9, and the currently observed population.}
	\label{tab:ecs}
	\begin{tabular}{ccccc}
		\hline
		Cases: & $m_9$ = $7.5 M_{\oplus}$ & $m_9$ = $15 M_{\oplus}$ & w/out P9  & Obs. \\
		\hline
		 $N_{ECs}(D> 10 km)$ & $1.8-3.6$ & $1-2$ & $0.7-1$  & 4 \\
		\hline
	\end{tabular}
	\centering
\end{table}

In the next section we analyze the orbital distribution of extreme Kuiper belt objects in order to understand the effects of Planet-9 on these objects and make predictions on their planetary architectures.

\subsection{The effect of Planet-9 on distant trans-Neptunian objects}

\begin{figure*}
\includegraphics[scale = 0.1]{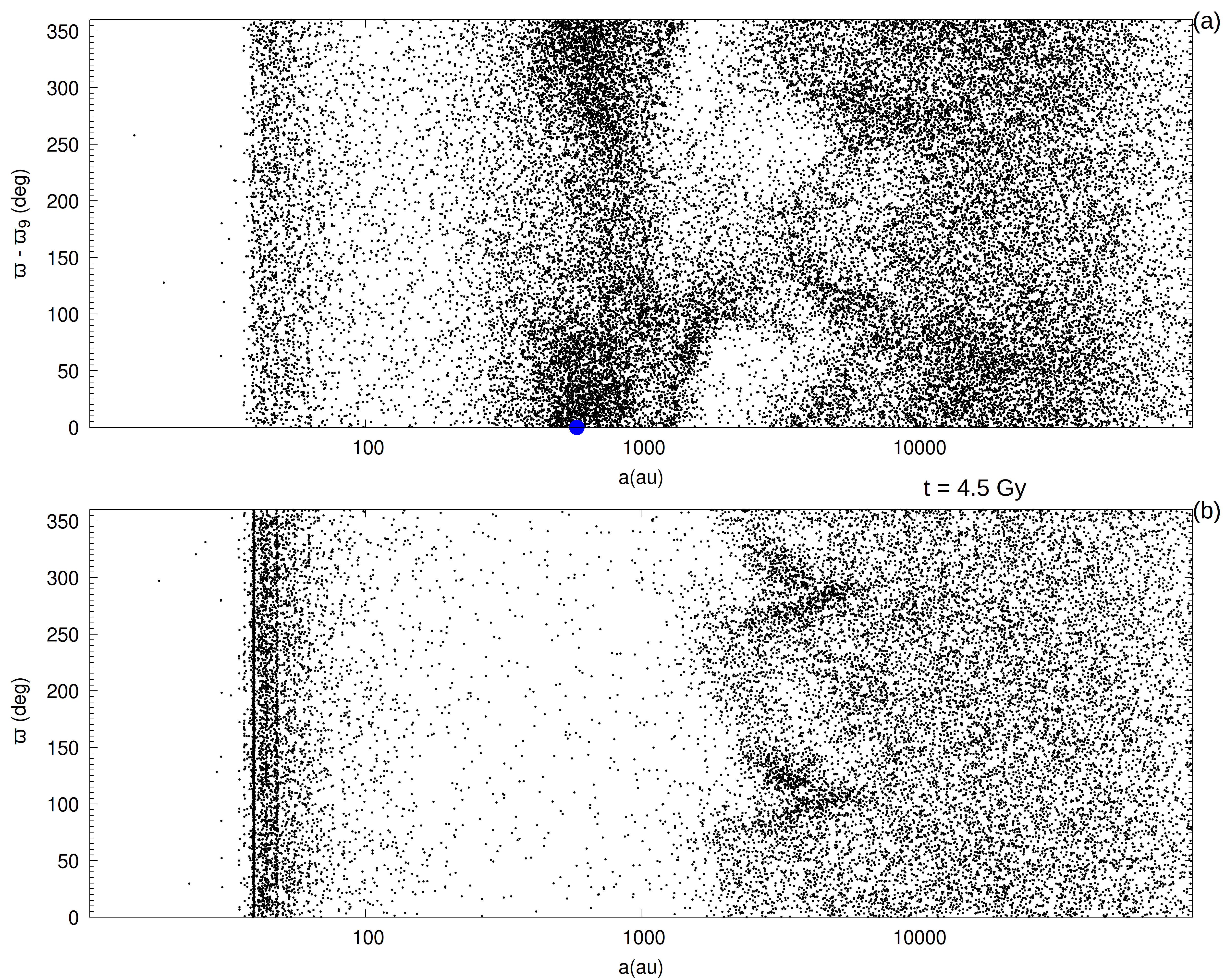}
\caption{Panel (a) shows a snapshot of our simulation at 4.5 Gyr depicting the relative perihelion longitudes ($\varpi - \varpi_9$) as function of  semi-major axis of all planetesimals. Panel (a) includes the effect our Planet-9. Panel (b) portrays the same snapshot but from a simulation where Planet-9 is not included. \label{fig:withp9}}.
\end{figure*}

As discussed earlier in this paper, several observed trans-Neptunian objects with  perihelion distance larger than $q > 40$ au and  semi-major axis greater than 150-200 au shows an unexpected orbital configuration characterized by an approximate clustering of the orbital angular momentum vectors and apsidal lines \citep{2014Natur.507..471T,2016AJ....151...22B,2021ApJ...910L..20B}. In this section, we analyze whether our simulation including the effects of Planet-9 produces such orbital clustering or not.


Figure \ref{fig:withp9}-a shows the final distribution of individual planetesimals in our simulation including the effects of Planet-9, at 4.5 Gyr. It shows the relative perihelion longitudes ($\varpi - \varpi_9$) as a function of semi-major axis.  Figure \ref{fig:withp9}-b shows the same distribution, but for planetesimals produced in a simulation without the presence of Planet 9, from \citet{N2017c}.

It is quite remarkable in Figure \ref{fig:withp9}-a that a significant fraction of planetesimals with semi-major axis between 200 and $\sim$1000 au are apsidally aligned with Planet-9, concentrating around 0. The blue dot in Figure \ref{fig:withp9}-a represents Planet-9, for reference. In contrast, no similar clustering is observed in Figure \ref{fig:withp9}-b which is a simulation without Planet-9 from \citet{N2017c}. Note that the prominent dynamical structures observed in Figures \ref{fig:withp9}-a (with Planet 9) and b (without Planet 9) between approximately 2000 and 10,000 au are the result of the influence of galactic tides \citep{higuchietal07}.

To further analyze the orbital alignment caused by Planet-9, we apply specific cutoffs to our population of simulated KBOs and plot histograms of the distribution of the longitude of perihelion of these objects relative to that of Planet-9.

We compare our results with observations by examining the ratio of apsidally anti-aligned and aligned objects using two slightly different approaches. First, we adopt the criteria from \citet[][]{2021ApJ...910L..20B}, defining apsidally anti-aligned objects as those with angular separations of $180^\circ \pm 90^\circ$, and apsidally aligned objects as those with angular separations of $0^\circ \pm 90^\circ$ at the end of our simulations. While these definitions are not mathematically rigorous in the context of resonant dynamics \citep[see, for instance,][]{BB2019}, they provide a practical framework for comparing our results with observations.

Based on this approach, among the 13 stable distant TNOs identified by \citet[][]{2021ApJ...910L..20B} ($40 < q < 100$~au and $i < 30^\circ$), 2 objects are apsidally aligned, and 11 are apsidally anti-aligned, yielding an observed ratio of approximately 5.5.

As a complementary analysis, we also applied a more restrictive definition of apsidally anti-aligned and aligned objects. In this secondary analysis, we define these populations as those with angular separations of $180^\circ \pm 40^\circ$ and $0^\circ \pm 40^\circ$, respectively. Using this stricter criterion, 2 of the observed TNOs are classified as apsidally aligned, and 7 are apsidally anti-aligned, while 4 objects fall outside both categories, yildeing a ratio of about 3.5. 

When comparing our results and observations, we crudely account for observational bias in our simulations by applying cutoffs to the orbital inclination and pericenter distance of the simulated planetesimal populations~\citep{BB2019}. Objects with low orbital inclinations and smaller pericenter distances are more likely to be detected due to these biases. Notably, all 13 stable trans-Neptunian objects (TNOs) identified by \citet[][]{2021ApJ...910L..20B} have orbital inclinations below approximately 30 degrees.

\begin{figure}
\centering
 \includegraphics[scale = 0.32]{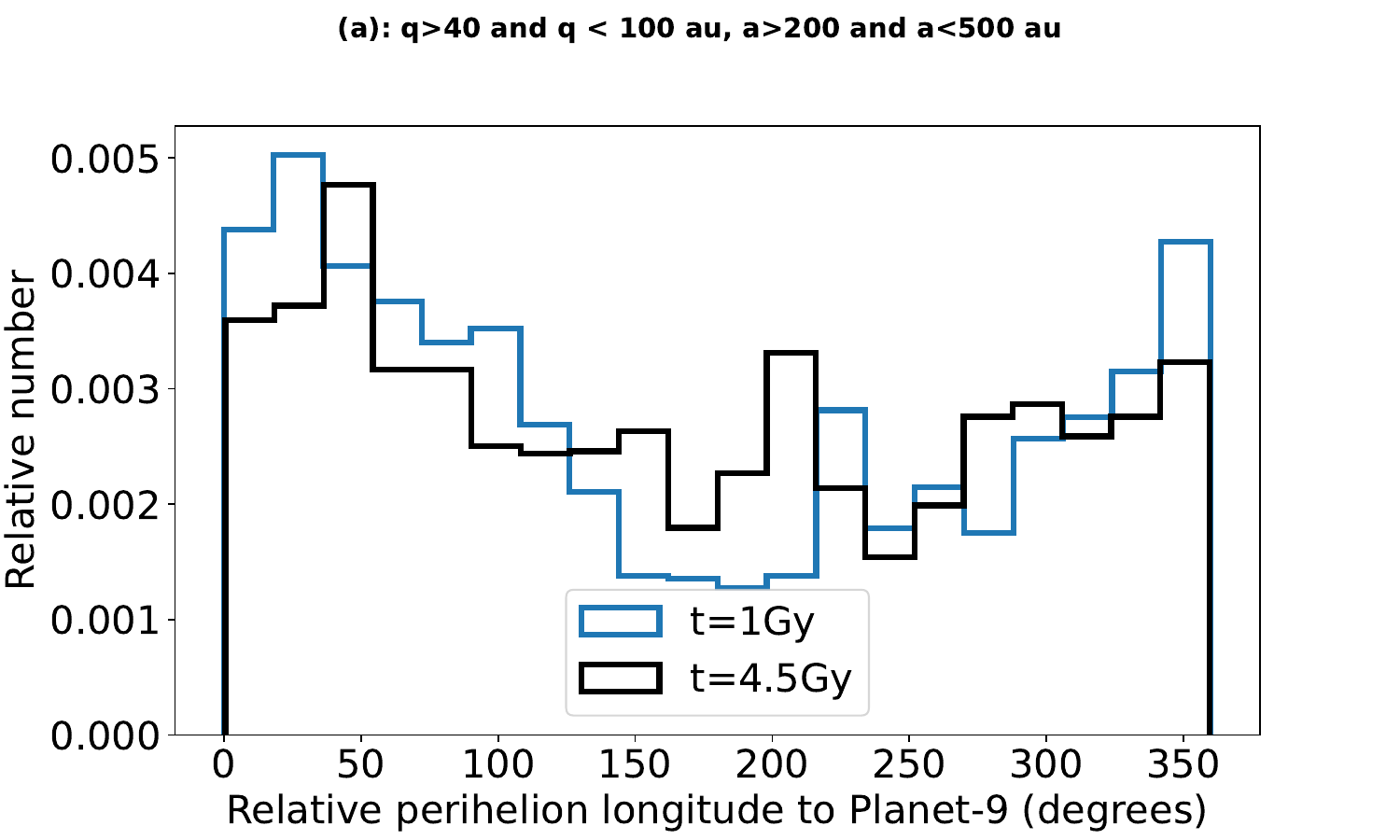}
 \includegraphics[scale = 0.32]{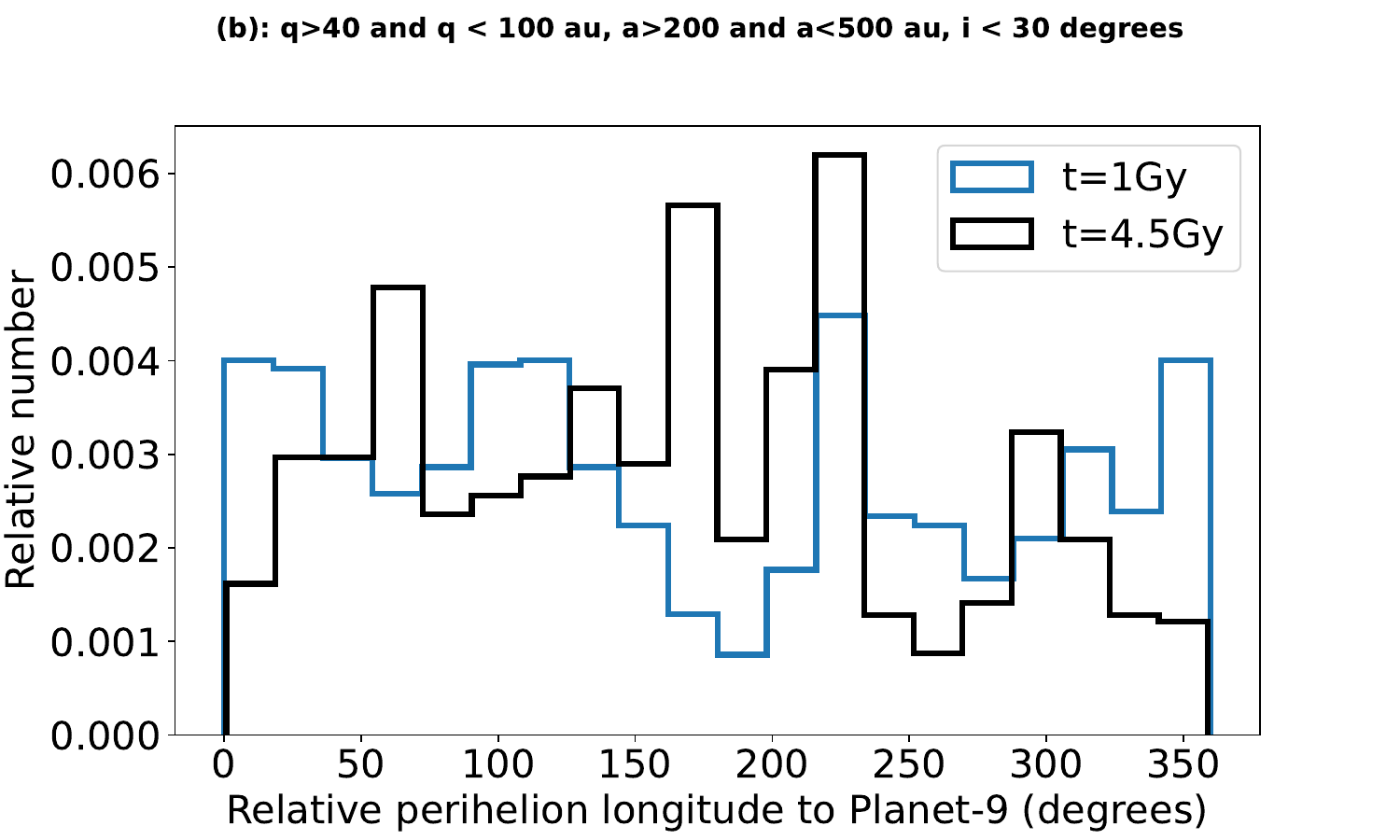}
 \includegraphics[scale = 0.32]{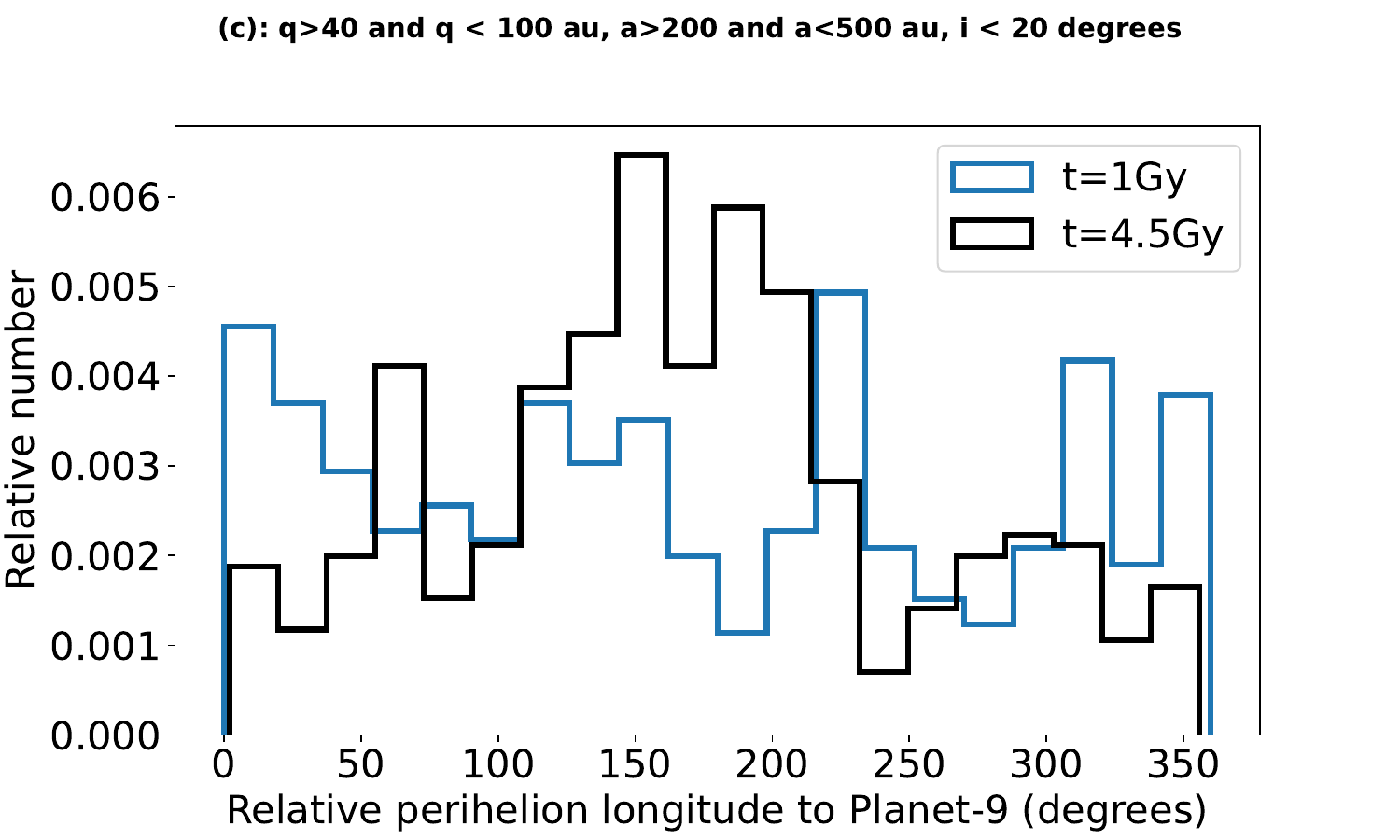}
 \includegraphics[scale = 0.32]{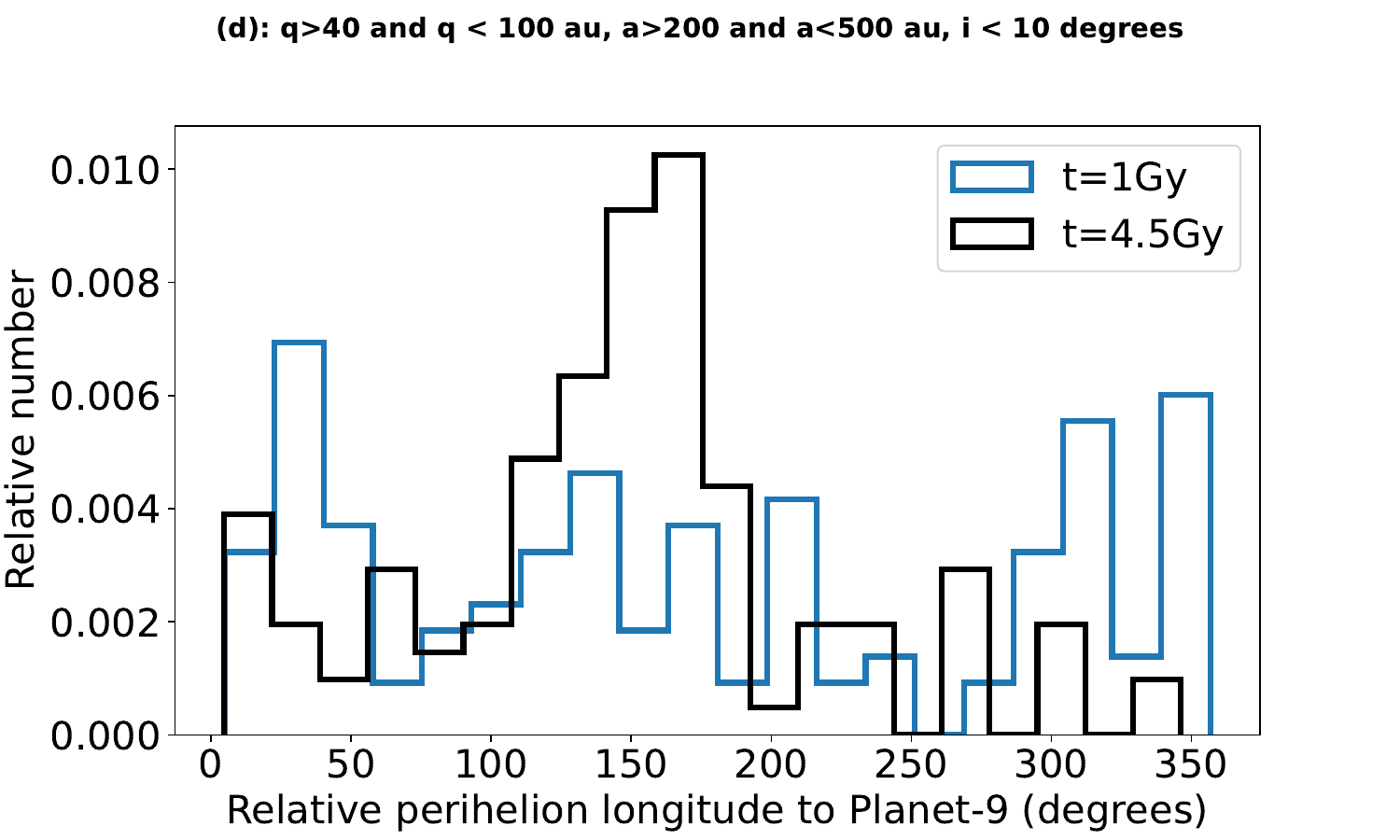}
\caption{Panels (a), (b), (c) and (d) show histograms depicting the relative number of simulated extreme KBOs with three different cuts: i) $q > 40$ au and $q < 100$ au  and $200< a < 500 $ au, ii) $q > 40$ au and $q < 100$ au  and $200< a < 500 $ au and $i < 30$ degrees, iii) $q > 40$ au and $q < 100$ au  and $200< a < 500 $ au and $i < 20$ degrees, and iv)  $q > 40$ au and $q < 100$ au  and $200< a < 500 $ au and $i < 10$ degrees, respectively. The blue and black lines represent two different times: 1 Gyr and 4.5 Gyr.\label{fig:newp2}}.
\end{figure}

Figure \ref{fig:newp2} shows the distributions of selected  distant Kuiper belt objects at 1 and 4.5 Gyr. In Figure \ref{fig:newp2}, we restrict our sample to objects with $40<q < 100$ au and $200< a < 500 $ au. Considered orbital cutoffs are indicated at the top of each panel.

The orbital distribution of objects in Figure \ref{fig:newp2}-a confirms the visual inspection of Figure \ref{fig:withp9},  showing a significant fraction of objects apsidally aligned with Planet-9. In Figure \ref{fig:newp2}-a, the number of objects aligned apsidally and anti-aligned is 1523 (662) and 1084 (515) , respectively, resulting in a anti-aligned-to-aligned ratio of 0.71 (0.8). The numbers given in parentheses come from our analysis considering our more restrictive definitions for apsidally anti-aligned and aligned objects. The total number of objects in Figure \ref{fig:newp2}-a is 2604.  We next further constrain our sample to objects with orbital inclinations lower than 30 degrees.

In Figure \ref{fig:newp2}-b, the number of apsidally aligned and anti-aligned objects is 366 (138) and 463 (273), giving a ratio of about 1.71 (1.98). The total number of objects in this case is 829. In Figure \ref{fig:newp2}-c, the number of aligned and anti-aligned objects is 167 (51) and 315 (204), respectively, giving a ratio of about 1.9 (3.94) (the total number of objects is 482). Finally, in Figure \ref{fig:newp2}-d these numbers are 39 (14) and 88 (54), yielding a ratio of approximately 2.25 (3.85). The total number of objects in Figure \ref{fig:newp2}-d is 121.

In terms of pericenter distance, most observed objects reported in \citet[][]{2021ApJ...910L..20B} have pericenter smaller than 50~au. That being said, we apply additional cuts in pericenter distance and recomputed the number of apsidally anti-aligned and aligned objects in each sub-sample.

The total number of objects with $40<q < 50$ au and $200< a < 500 $ au in our simulations is 585, with  191 (111) being apsidally anti-aligned and 394 (170) aligned, respectively. This results in a ratio of about 0.48 (0.65). If we apply further cutoffs in inclination as in Figure \ref{fig:newp2}, the number of objects with $i<30$ deg becomes  (225), where 115  (69) are apsidally anti-aligned and 110 (30) are aligned, respectively, resulting in a ratio of 1.05 (2.3). If we further restrict our sample to objects with $i<20$ deg, we are left with 164 objects, where 120 (59) are anti-aligned and 44 (19) are aligned. The ratio in this case is 2.72 (3.1)  Finally, if we account only for objects with $i<10$ deg, it remains only 23 objects, where 6 (13) are anti-aligned and 10 (0) objects are aligned.

Motivated by the findings of \cite{batyginetal24}, we also analyzed the number of low-inclination, Neptune-crossing objects (Centaurs) in our simulations with and without the presence of Planet-9. \cite{batyginetal24} propose that if Planet-9 exists, it should consistently produce nearly planar ($i < 40^\circ$), long-period ($100 \, \mathrm{AU} < a \lesssim 1000 \, \mathrm{AU}$) objects with perihelia smaller than $q < 30 \, \mathrm{AU}$. This hypothesis is potentially supported by the observation of more than a dozen objects with these orbital properties. Our objective here is not to directly compare our results with observations, but rather with those of the simulations of \cite{batyginetal24}, which follow a different approach from that used in this work.

\cite{batyginetal24} report a ratio of objects with Neptune-crossing orbits ($q < 30$~au), inclinations $i < 40^\circ$, and semi-major axes between 100 and 1000~au to those with $q > 30$~au (also $100<a<1000$~au; $i<40$) of 3\% in simulations including Planet 9 and 0.5\% in simulations without it, representing a factor of 6 difference. In our simulations, we found a ratio of 2.6\% in scenarios with Planet 9 and 1.4\% without, resulting in a factor of approximately 2. When we extend the semi-major axis range to 100--2000~au, our results change to 2.37\% and 0.77\% for simulations with and without Planet 9, respectively. We suggest that this broader range of semi-major axes, compared to that considered by \cite{batyginetal24}, provides a more comprehensive representation of objects influenced by Planet 9 in our simulations (see our Figure \ref{fig:case_a_e}). From this analysis, the ratio of objects produced in our simulations is a factor of 2 or 3 lower than that reported by \cite{batyginetal24}. The origin of these differences remains unclear. One possible explanation, particularly in the context of Planet 9, could be the differing initial conditions assumed for Planet 9 in the simulations conducted by \cite{batyginetal24} compared to those used in this work. In \cite{batyginetal24}, Planet 9 is assumed to have a mass of 5 $M_{\oplus}$, semi-major axis of $a=500$~au, eccentricity of $e=0.25$, and orbital inclination equal to $i=20$~deg. In our case, Planet 9 is assumed to have a mass of 7.5 $M_{\oplus}$, $a=600$~au, $e=0.3$, and $i=20$~deg. Conversely, the observed differences in cases without Planet 9 may be attributed to Neptune's migration, a process that was not accounted for in  \cite{batyginetal24}. Neglecting this process may lead to different initial conditions for the TNO population compared to our work.

We conclude this section by suggesting that the majority of trans-Neptunian objects influenced by Planet 9 are likely to be predominantly apsidally aligned with this hypothetical planet. Apsidally anti-aligned objects outnumber apsidally aligned ones only within sub-populations characterized by relatively low orbital inclinations ($\lesssim$10-20$^\circ$; comparable to the assumed inclination of Planet-9) and small perihelion distances ($q \lesssim 100$ au). Assuming these cutoffs provide a reasonable approach to including observational bias in our simulation \citep{2021ApJ...910L..20B}, our results indicate that the  ratio of the number of objects  apsidally anti-aligned to aligned in the {\it current observed population} (estimated as $\sim$3.5-5.5) is reasonably close to that produced in our simulation including Planet 9 ($\sim$2-3.9). We note, however, that the observed population is limited by small-number statistics, and comparisons between simulations and observations should be interpreted with caution.

\section{Conclusions}
\label{conclusion}

In this work, we modelled the the formation and evolution of the  trans-Neptunian object population and Oort cloud during the migration phase of the giant planets, and subsequent solar system evolution, for 4.5 Gyr. We compared the distributions of objects in these different reservoirs --  in simulations including and neglecting the effects of Planet-9 --  with observations. In many aspects, this work is similar to that of \citet{N2017c}, with an important  difference being that in our simulations we have updated the mass of Planet-9, as recently constrained \citep{BB2019}.




The main conclusions of this work are as follows: i) The orbital distribution, including the inclination distribution, of ecliptic comets is well reproduced in our model with a Planet-9 having a mass of $m_9 = 7.5 M_{\oplus}$,
an inclination of $i_9 = 20$ degrees, a semi-major axis of $a_9 \sim 600$ au, and an eccentricity of $e_9 \sim 0.3$. This result is a significant improvement in the model compared to those of  \citet{N2017c}, where the inclination distribution of ecliptic comets was more spread than the observed one in simulations assuming a mass for  Planet-9 twice higher than that used in our work. ii) Our nominal model predicts a number of ecliptic comets with $D>10$~Km ranging between $1.8-3.6$. Current observations suggest the existence of about 4 ecliptic comets with $D>10$~km, suggesting a good match between our model and observations, although it is not entirely clear how complete observations are.

Finally, our simulations predict that the majority of  distant Kuiper belt objects are apsidally aligned with Planet-9, if it exists. Only sub-populations of objects with orbital inclination lower than $\lesssim$20 deg (comparable to that of Planet-9) and pericenter smaller than 100~au are expected to show a relatively significant fraction of apsidally anti-aligned objects. These results can help to guide our search to find or to rule out the existence of Planet-9.

\section*{Acknowledgements \label{Sec:Acknowledgements}}

We are very thankful to Julio Fernandez and an anonymous referee for their constructive comments, which significantly improved an earlier version of this manuscript. Rafael Ribeiro (RR) thanks to the scholarship granted from the Brazilian Federal Agency for Support and Evaluation of Graduate Education (CAPES), in the scope of the Program CAPES-PrInt (Proc~88887.310463/2018-00, Mobility number 88887.572647/2020-00, 88887.468205/2019-00). This research was supported in part by the São Paulo Research Foundation (FAPESP) through the computational resources provided by the Center for Scientific Computing (NCC/GridUNESP) of the São Paulo State University (UNESP). RR also  acknowledges the support provided by grants FAPESP (Proc~2016/24561-0) and by Sao Paulo State University (PROPe~13/2022).   A.~I. is thankful to the NASA Emerging Worlds Program for financial support (Grant n. 80NSSC23K0868).

\clearpage
\appendix
\setcounter{figure}{0}
\renewcommand{\theHfigure}{A\arabic{figure}}
\renewcommand{\thefigure}{A\arabic{figure}}

\section{Additional Results}

In this Section we present additional plots that support and complement the discussion and results presented in our main paper.



\begin{figure}
   \centering
 \includegraphics[scale = 0.32]{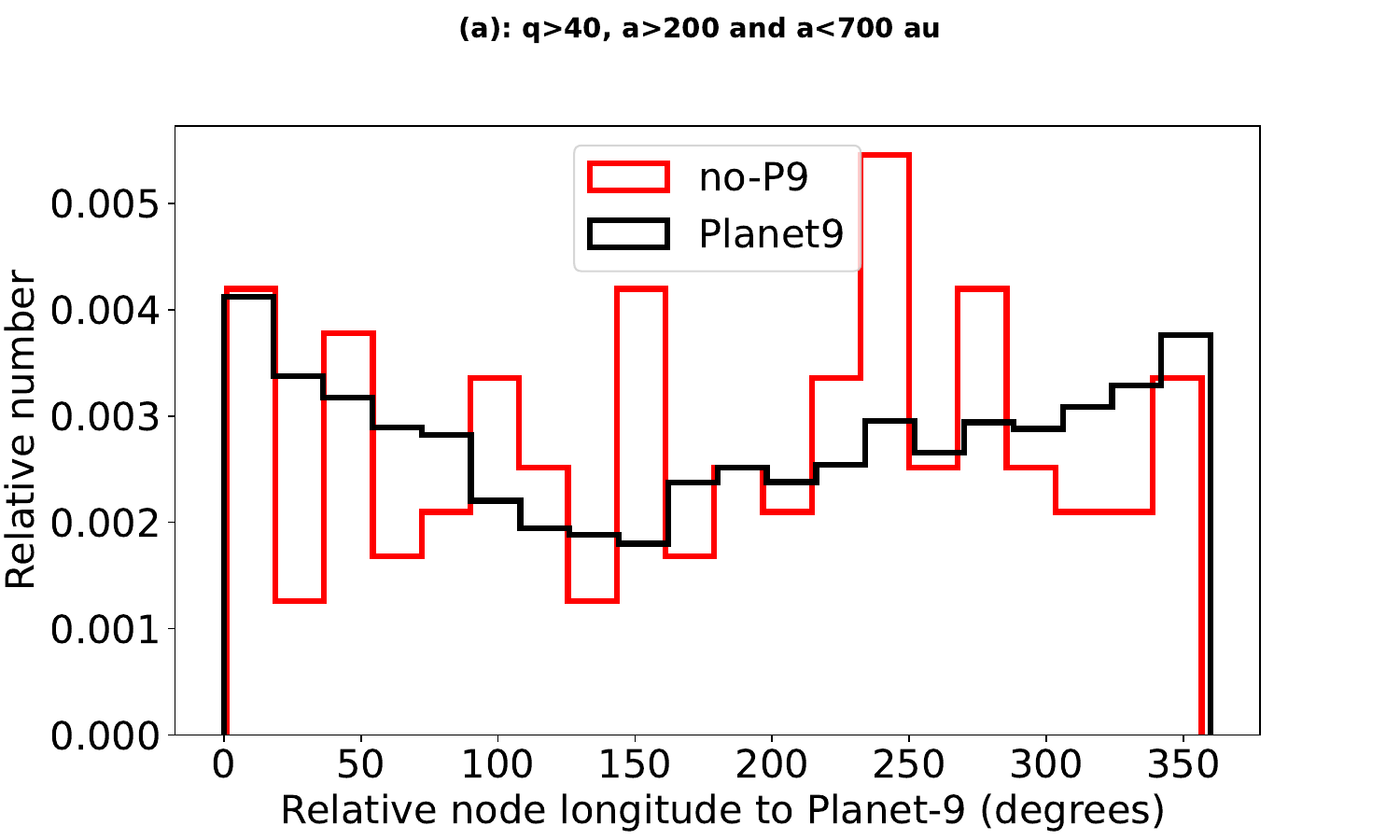}
 \includegraphics[scale = 0.32]{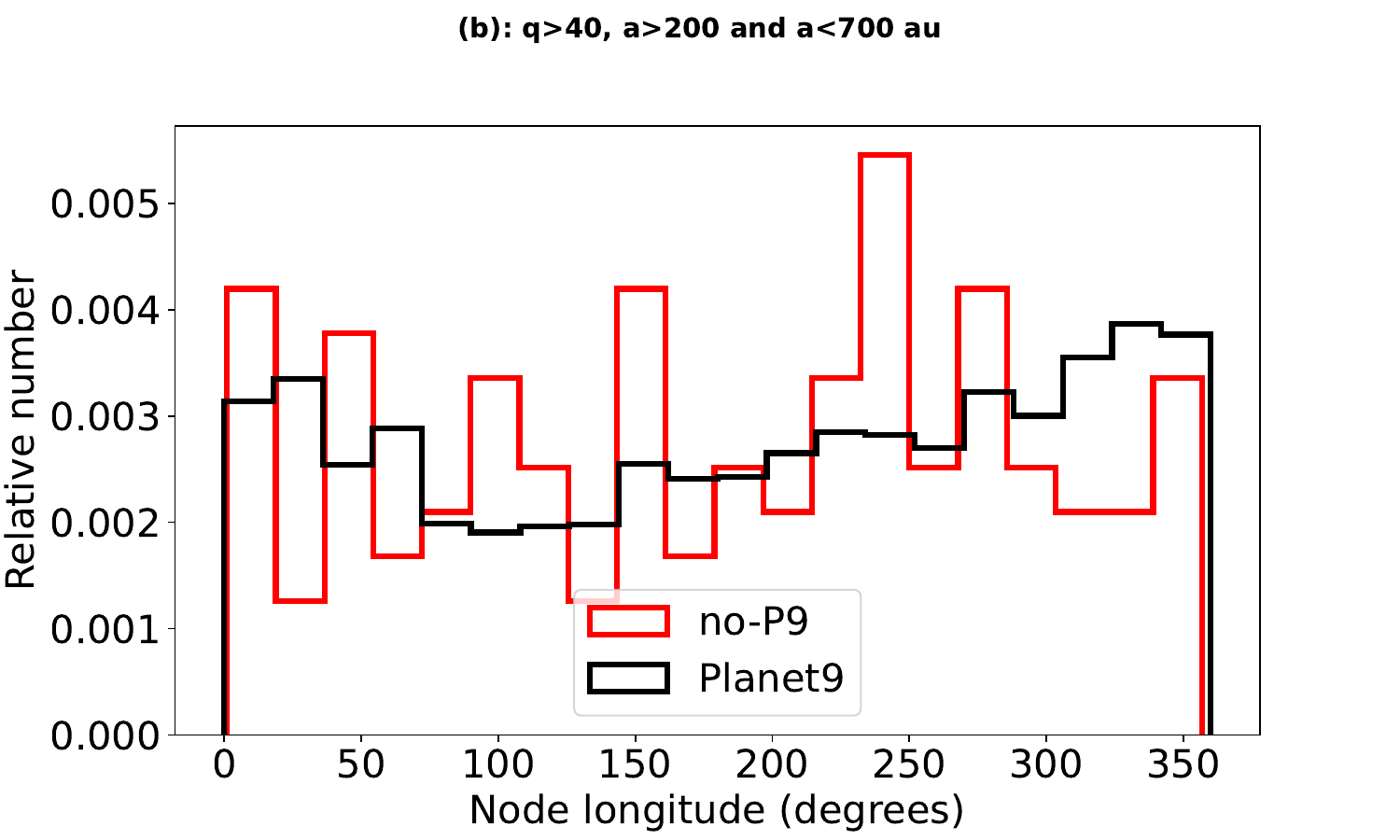}
 \includegraphics[scale = 0.32]{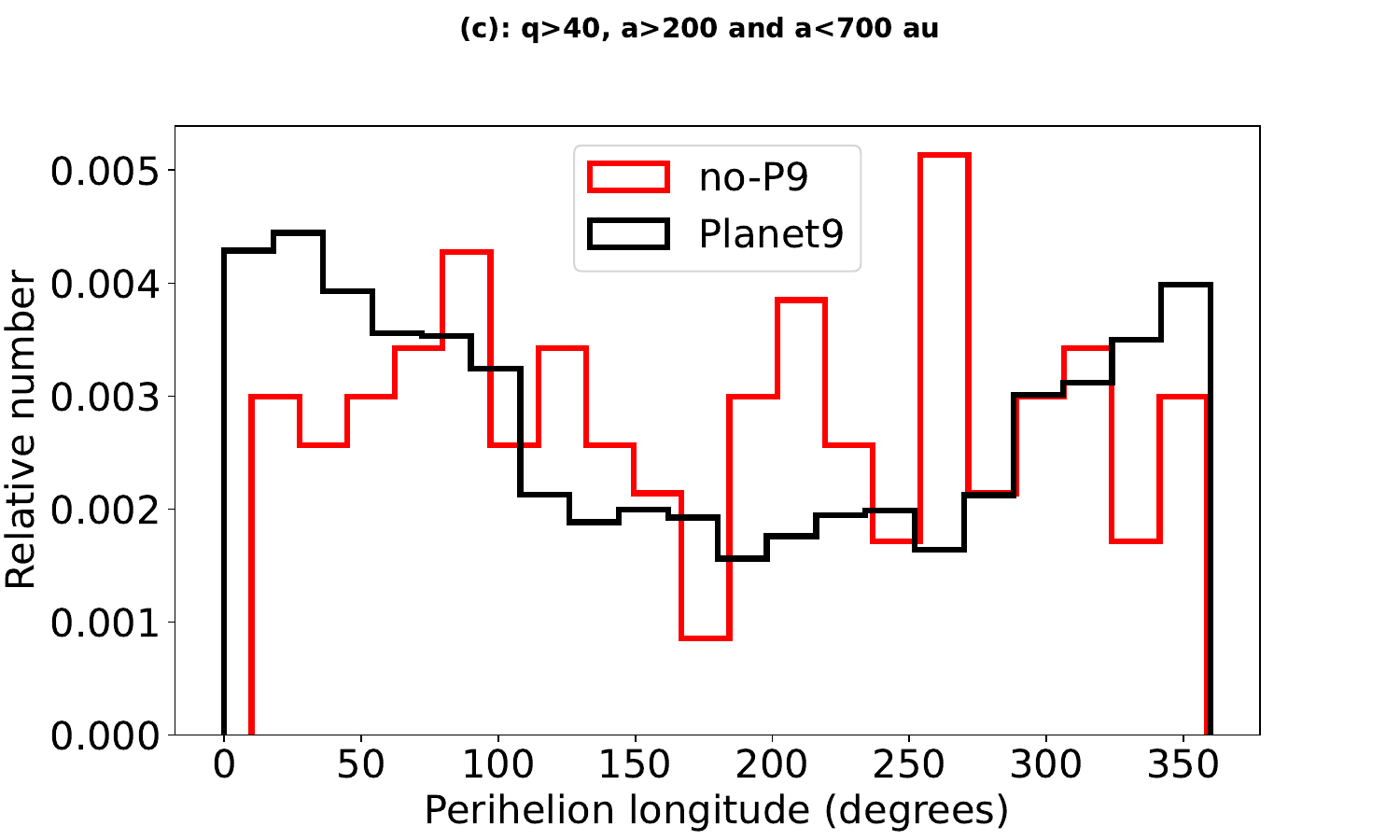}
  \includegraphics[scale = 0.32]{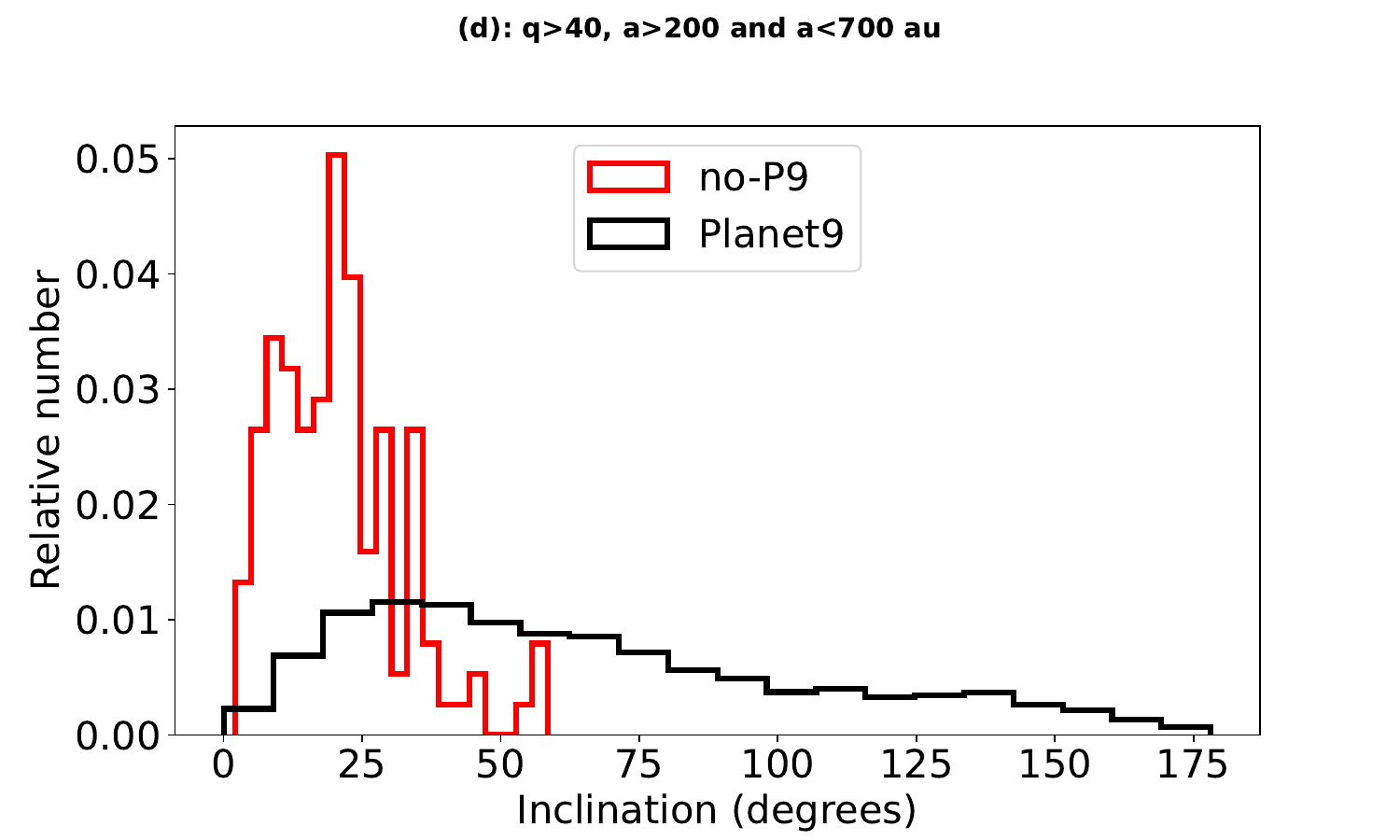}
\caption{Orbital distribution of simulated distant Kuiper-belt objects in simulations including (black) and neglecting the effects of Planet-9 (red). Orbital cutoffs applied to the original sample is shown at the top of each panel.\label{fig:hist2}}.
\end{figure}

We complement our analysis by showing in Figure \ref{fig:hist2}  the distributions of the relative node longitude to Planet-9, node longitude, perihelion longitude, and  orbital inclination of selected distant KBOs, as indicate at the top of each panel. The red histograms show the case without Planet-9 from \citet{N2017c}, and black histograms show the results of our simulation. Figure \ref{fig:hist2} a) and b) show that our selected population of distant KBOs is mostly aligned with Planet-9 also in terms of their longitude of the node. The distributions of longitudes of node and perihelion also points to orbital alignment (Figure \ref{fig:hist2}-c). On the contrary, Figure \ref{fig:hist2} also shows that no clear orbital clustering exists in simulations without Planet-9 (red histograms).

Figure \ref{fig:hist2} also shows the orbital inclination distribution of selected KBOs in our simulations (with $q>40$~au and $200<a<700$~au) including and without Planet-9. Overall, the orbital inclination distribution produced in our simulation with Planet-9 is much more spread than that of the simulation without Planet-9. In our simulation, the orbital inclination peaks at about 40 degrees. In the case without Planet-9, it is typically lower than 25 degrees.

\begin{figure*}
\centering

\includegraphics[scale = 0.32]{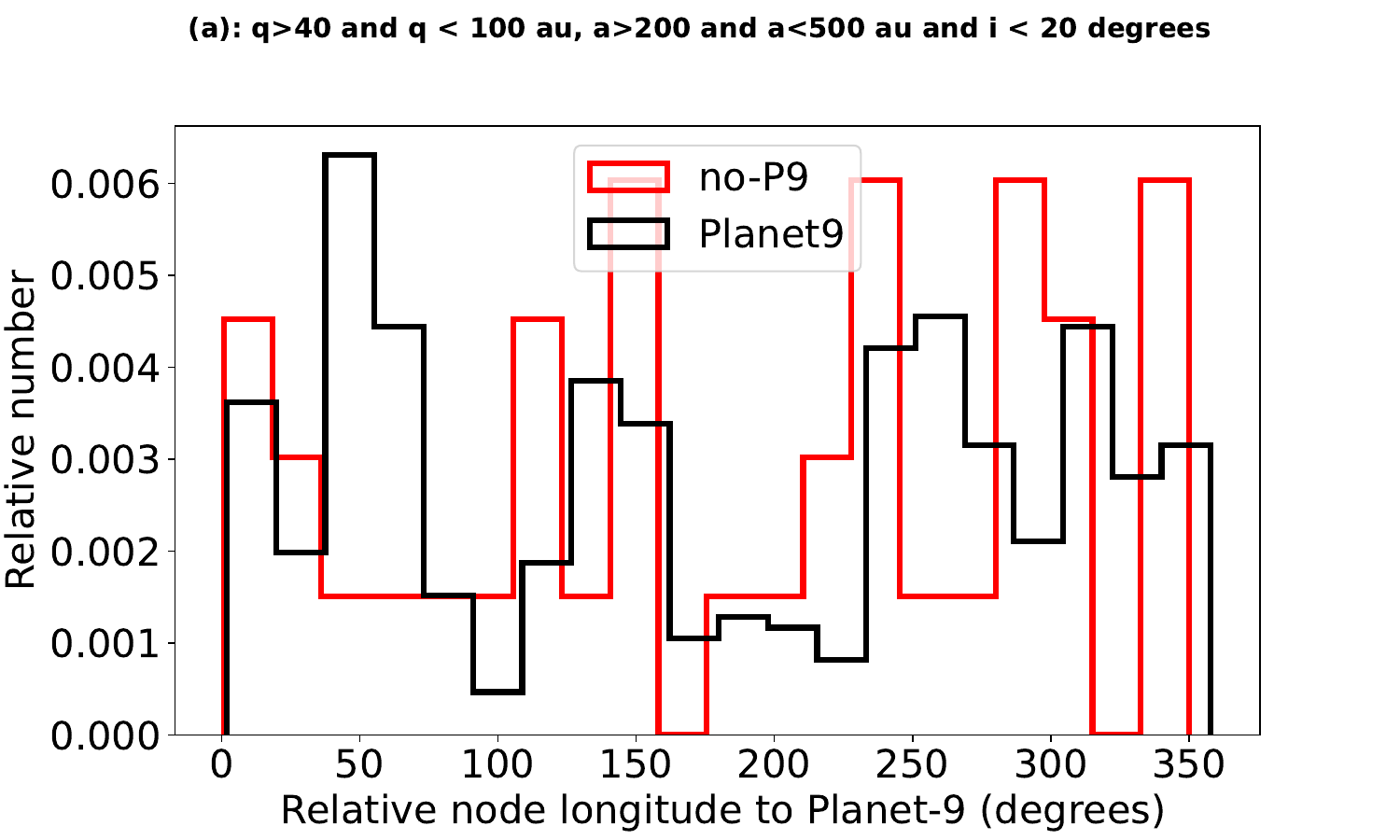}
\includegraphics[scale = 0.32]{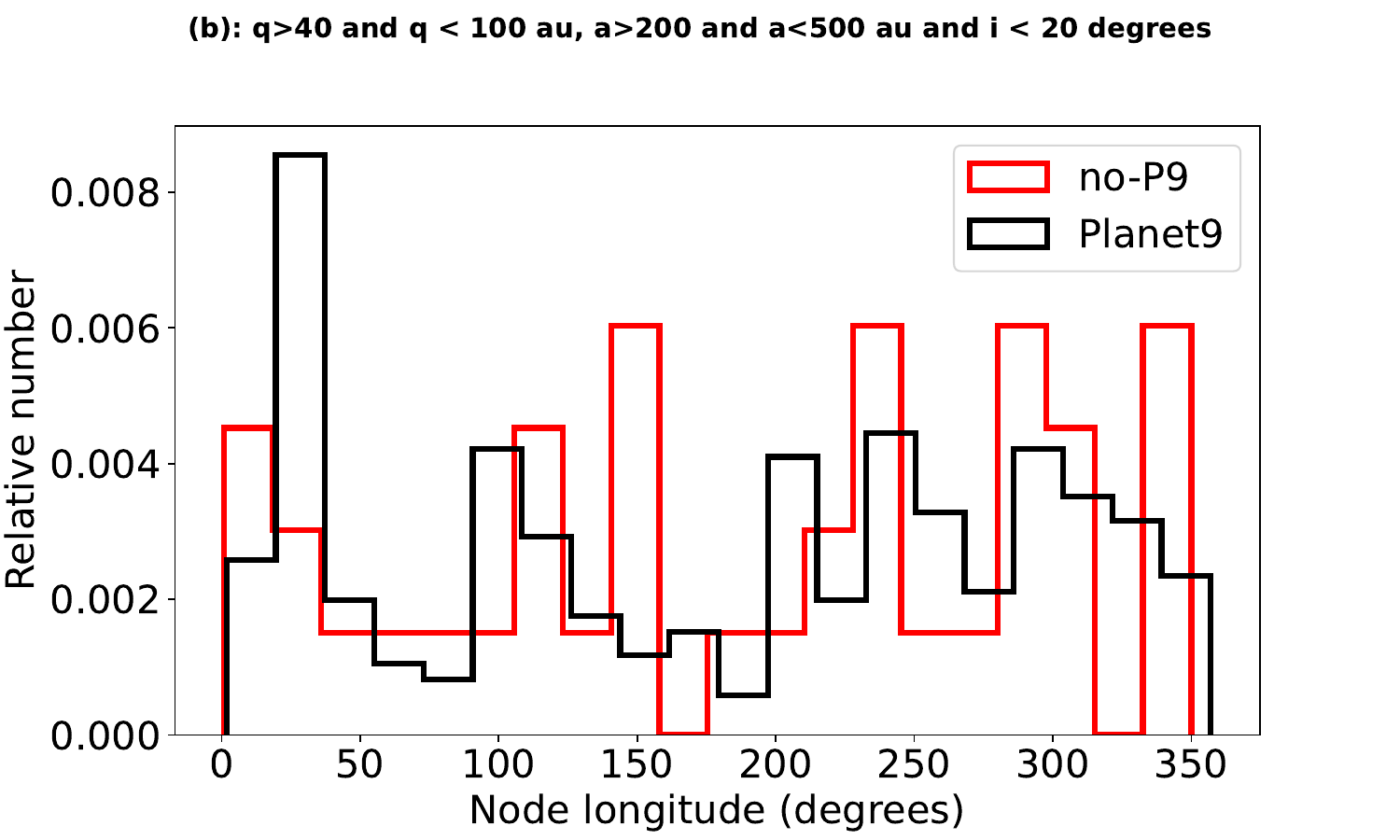}
\includegraphics[scale = 0.32]{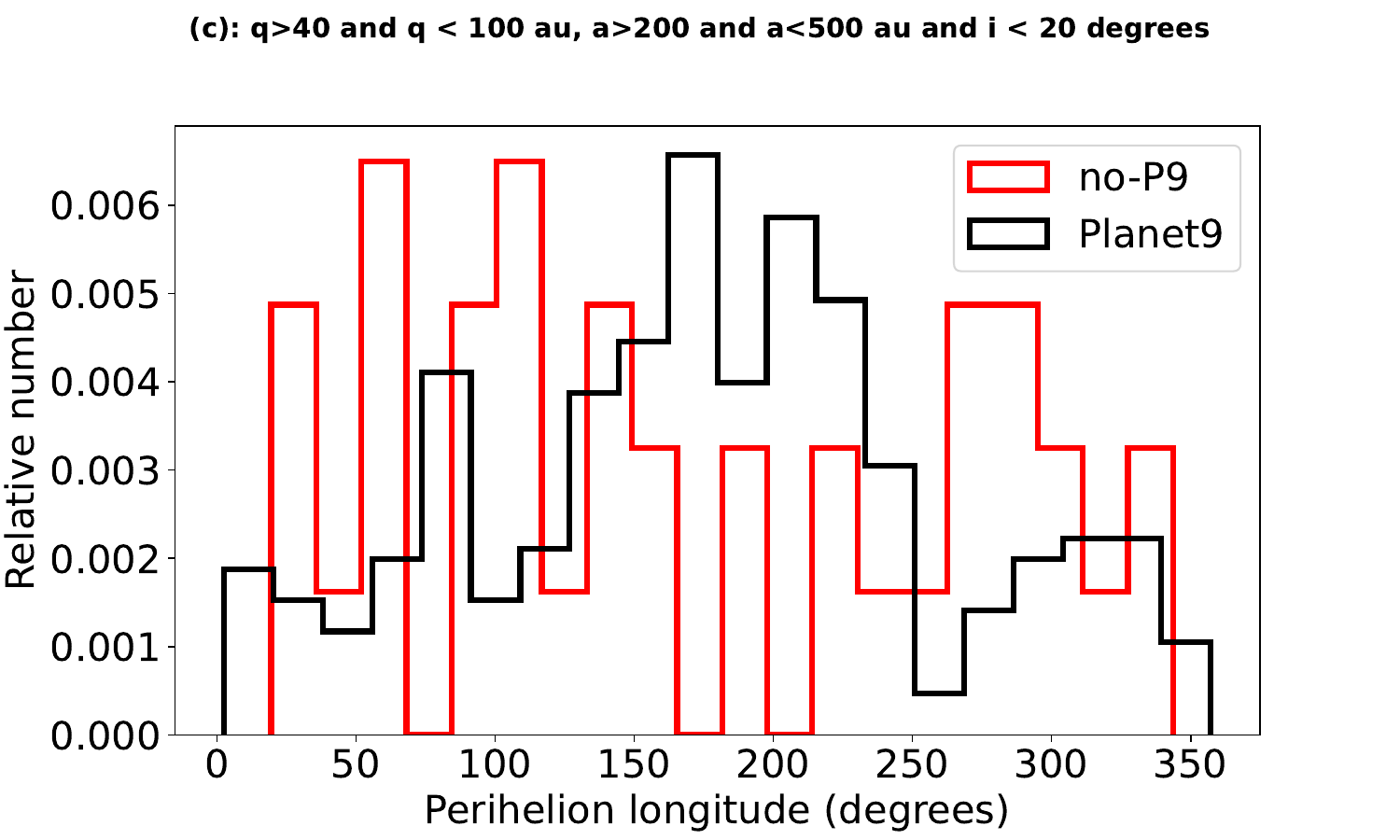}
\includegraphics[scale = 0.32]{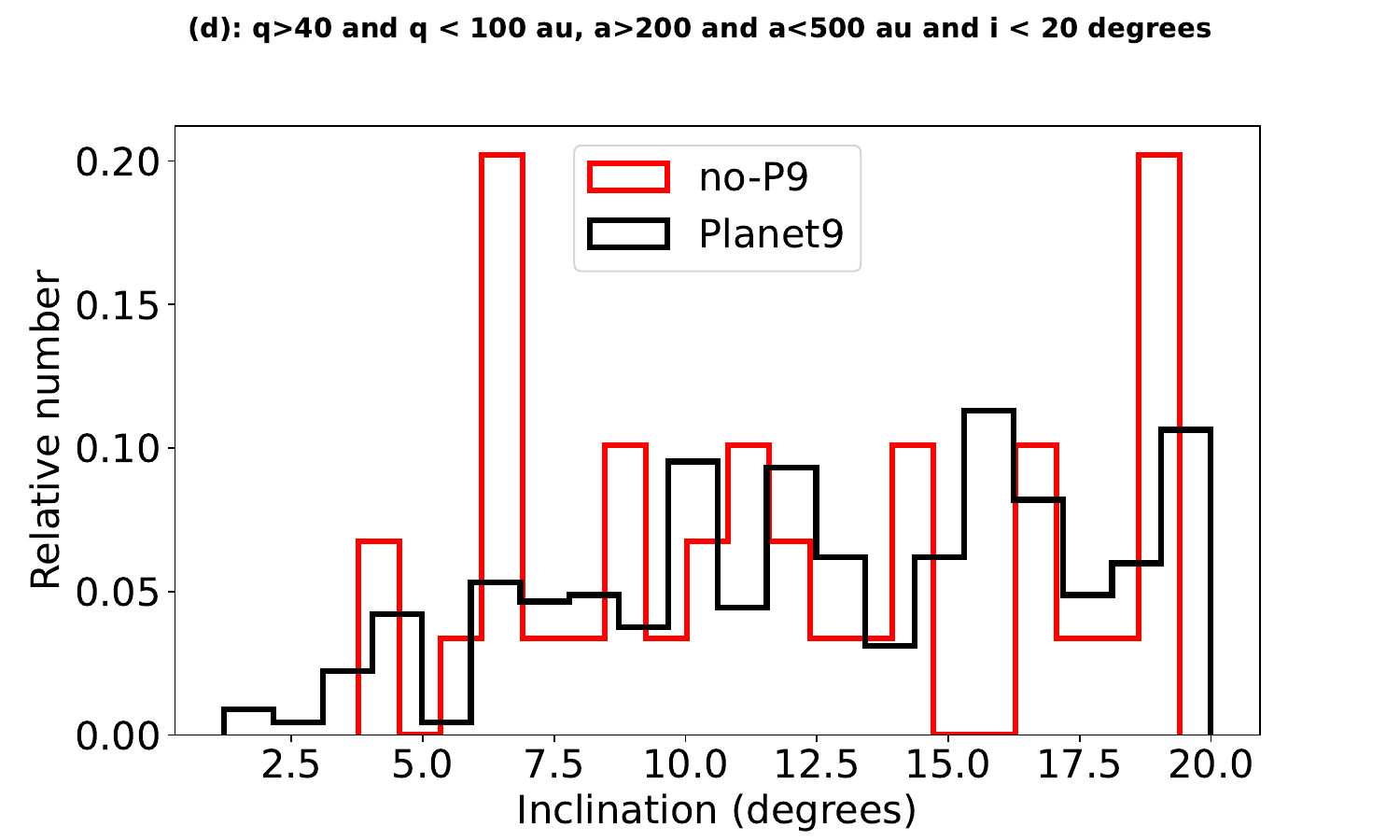}
\caption{Orbital distribution of simulated distant Kuiper-belt objects in simulations including (black) and neglecting the effects of Planet-9 (red). Orbital cutoffs applied to the original sample are shown at the top of each panel.\label{fig:hist4}}.
\end{figure*}

Figure \ref{fig:hist4} shows the distributions as in Figure \ref{fig:hist2}, but accounts only for objects with orbital inclinations lower than 20 degrees and $q<100$~au. The relative node longitude to Planet-9 shows some level of orbital clustering around 0. We have calculated the ratio between the number of anti-aligned and aligned objects (as defined as those with angular separations of $180^\circ \pm 40^\circ$ and $0^\circ \pm 40^\circ$, respectively) as 1.6. We do not show histograms for objects with orbital inclinations lower than 10 degrees due to the small number of objects in the sub-sample. For objects with $40<q<100$~au, $200<a<500$ and $i<10$ degrees, the ratio between the number of anti-aligned and aligned objects for the relative node longitude is about 3.




\label{lastpage}



%
\bibliographystyle{aasjournal}

\bibliography{ribeiroetal2023} 

\end{document}